\documentclass[prb,twocolumn,showpacs,10pt]{revtex4}
\usepackage[dvips]{graphicx}
\usepackage{color,array,dcolumn}
\usepackage{amsmath}
\usepackage{amssymb}

\usepackage{amsmath}
\usepackage{amssymb}
\usepackage{mathrsfs}
\allowdisplaybreaks{}
\begin{document}

\title{Hole pairing from attraction of opposite chirality spin vortices:\\
Non-BCS superconductivity in Underdoped Cuprates}

\author{P. A. Marchetti}

\affiliation{Dipartimento di Fisica, INFN, I-35131 Padova, Italy}

\author{F. Ye}

\affiliation{College of Material Sciences and Optoelectric Technology,
  Graduate University of Chinese Academy of Science, Beijing 100049,
  China}

\author{Z. B. Su}

\affiliation{Institute of Theoretical Physics, Chinese Academy of
  Sciences, 100190 Beijing, China}

\author{L. Yu}

\affiliation{Beijing National Laboratory for Condensed Matter Physics
  and Institute of Physics, Chinese Academy of Sciences, 100190 Beijing,
  China, }

\affiliation{Institute of Theoretical Physics, Chinese Academy of
  Sciences, 100190 Beijing, China}

\begin{abstract}
  Within a gauge approach to the $t-J$ model, we propose a new, non-BCS
  mechanism of superconductivity (SC) for underdoped cuprates.  We
  implement the no-double occupancy constraint with a (semionic)
  slave-particle formalism. The dopant in the $t-J$ model description
  generates a vortex-like quantum distortion of the antiferromagnetic
  (AF) background centered on the empty sites, with opposite chirality
  for cores on the two N\'eel sublattices. Empty sites are described in
  terms of spinless fermionic holons and the long-range attraction
  between spin vortices on two opposite N\'eel sublattices serves as the
  holon pairing force, leading eventually to SC. The spin fluctuations
  are described by bosonic spinons with a gap generated by scattering on
  spin vortices. Due to the no-double occupation constraint, there is a
  gauge attraction between holon and spinon, binding them into a
  physical hole. Through gauge interaction the spin vortex attraction
  induces the formation of spin-singlet (RVB) spin pairs by lowering the
  spinon gap, due to the appearance of spin-vortex dipoles.  Lowering
  the temperature, the proposed approach anticipates two crossover
  temperatures as precursors of the SC transition: at the higher
  crossover a finite density of incoherent holon pairs are formed,
  leading to reduction of the hole spectral weight, while at the lower
  crossover a finite density of incoherent spinon RVB pairs are also
  formed, giving rise to a gas of incoherent preformed hole pairs with
  magnetic vortices appearing in the plasma phase, supporting a Nernst
  signal. Finally, at an even lower temperature the hole pairs become
  coherent, the magnetic vortices become dilute and SC appears beyond a
  critical doping.  The proposed SC mechanism is not of the BCS-type,
  because it involves a gain in kinetic energy, due to the lowering of
  spinon gap, and it is ``almost'' of the  classical 3D
  XY-type. Since both
  the spinon gap, describing short-range AF order, and the holon
  pairing, generating SC, originate from the same term in the
  slave-particle representation of the $t-J$ model, the proposed
  approach incorporates a strong interplay between AF and SC, giving
  rise to a universal relation between $T_c$ and the energy of the
  resonance mode, as observed in neutron scattering experiments.
\end{abstract}
\maketitle
\section{Introduction}

The high temperature superconductivity (SC) in cuprates,
discovered 25 years ago,\cite{muller} still remains a major
challenge in the condensed matter physics. In spite of the
enormous progress made in materials synthesis, crystal growth,
experimental studies of physical properties and theoretical
interpretation, there is still no consensus yet regarding the
anomalous normal-state properties and SC mechanisms in these
cuprate compounds.
 There is a recent
  review article\cite{lee2008} on various approaches attacking this
  extremely difficult problem, including Resonance Valence Bond (RVB)
  slave particle gauge approaches, spin fluctuation models, stripe
  models, phonons, three-band scenario, {\it etc.} We share many
  viewpoints expressed there, and
% Various approaches attacking this extremely hard problem, including the
% Resonance Valence Bond (RVB) slave particle gauge approaches, spin
% fluctuation models, stripe models, phonons, three-band scenario, etc,
% have been summarized in a recent review article,\cite{lee2008} and we
% share many viewpoints expressed there.
to save space, we refer the readers to that review article, not
repeating those comments here.

In this paper we propose a new mechanism of SC in hole-underdoped High
T$_c$ cuprates, using the spin--charge gauge approach to the 2D {\it
  t-J} model, describing the Cu-O planes.\cite{marchetti2007} In this
approach the {\it t-J} model (with {\it t/J} as the main
parameter) satisfying the single-occupancy constraint, is treated
systematically within the same set of approximations, to study
both normal-state and SC properties. The exchange $J$-term giving
rise to antiferromagnetism (AF) is also serving as the ``glue''
leading to SC, thus implementing the interplay of AF and SC in an
explicit form. The proposed SC mechanism is not of the BCS-type,
and it involves a gain in kinetic energy by lowering the spinon
gap due to the appearance of spin-vortex dipoles. The main
features of this non-BCS description of SC are consistent with the
experimental results in underdoped cuprates,
 including a natural $d$-wave SC pairing parameter,
and especially the contour plot of the Nernst
signal.\cite{ong1,ong2} We can also derive the SC transition as
``almost'' of the classical 3D XY-type, while the calculated
transition temperature shows a universal ratio to the resonance
mode energy observed in neutron experiments.\cite{Yu2009}

Our formalism basically belongs to the ``strong correlation--slave
particle tent'', where, a U(1) field is introduced to gauge the
global charge, while a SU(2) field is introduced to gauge the
global spin. Through the gauge field, a vortex-like quantum
distortion of the AF background is generated around the empty site
(described in terms of fermionic spinless holon) with opposite
chirality for cores on two N\'eel sublattices. In the presence of
these vortices the spin excitations (bosonic spin 1/2 spinon),
originally gapless without doping, corresponding to a long range
AF order, acquire a finite gap due to scattering on these vortices
(similar to the wave localization of light propagating in random
media), converting LR AFO to a short range order (SRO). Within
this approach the physical hole is a bound state of holon and
spinon with a ``glue'' (binding force)  coming from an
 emergent
U(1) slave particle gauge field. Here the spinon and holon are
neither confined (as in the ordinary Fermi liquid), nor decoupled
(as in 1D {\it t--J} model in the small $J$ limit), but rather
forming a ``composite particle''--physical hole in a strongly
correlated system.  It is not anymore a ``neat'' quasiparticle,
but rather having a strongly temperature-dependent lifetime due to
the gauge field (coupled to holons with a finite Fermi surface).
Similarly, the magnon is a composite particle made of spinon and
antispinon, again with the ``gauge glue''. In fact, the
``composite'' characteristics are responsible for all exotic
properties in the ``pseudogap phase''(PG), in particular, the
interplay of the SR AFO (exhibited as a finite magnon mass gap)
with the dissipative motion of charge carriers, showing up as
lifetime effect of the physical hole, results in a metal-insulator
crossover, a pronounced phenomenon in the underdoped cuprates. A
number of peculiar features of cuprates in the normal state can be
well explained within this scheme.\cite{marchetti2007} Here this
approach is generalized to consider the SC state.

The gluing force of the SC mechanism is an attraction between holons
generated by spin vortices on two opposite N\'eel sublattices, centered
around the empty sites (holes). This attraction which shares the same
origin of spin exchange $J$-term leading to AFO, was neglected as a
subleading term in considering the normal state properties.  Physically,
the hole is assigned an additional  ``pseudospin'' index marking
the belonging N\'eel sublattice in a range characterized typically by
the AF correlation length, as long as the SR AFO persists.  This
attraction describes the tendency towards vortex-antivortex binding, or
reduction of the AF exchange energy loss. In fact, the formation of the
vortex dipoles effectively reduces the density of free vortices
scattering the spin waves, and as a consequence the kinetic energy of
spinon gains and the AF correlation is enhanced as well.  To materialize
the SC transition we propose the following three-step scenario:

At the highest crossover temperature, denoted as $T_{ph}$, a
finite density of incoherent holon pairs are formed.  We propose
to identify that temperature with the experimentally observed
(upper) PG temperature, where the in-plane resistivity deviates
from the linear behavior.  A BCS-like $d$-wave pairing of holons
is derived by ``superposing'' two $p$-wave like pairing in a
reduced Brillouin zone. However, the holon pairing alone is not
enough for SC to appear. Again, through the ``gauge glue'' coming
from the U(1) slave particle gauge field, so crucial for the
interpretation of exotic properties of the PG phase, the spin
vortex attraction induces  the formation of spin-singlet (RVB)
spinon pairs with a reduction of the spinon gap. Physically, the
spinons will feel ``less disturbance'' from the formation of
vortex-antivortex pairs (dipoles).

At the intermediate crossover temperature, denoted as $T_{ps}$, a
finite density of incoherent spinon RVB pairs are formed, which,
combined with the holon pairs, gives rise to a gas of incoherent
preformed hole pairs. We propose to identify that temperature with
the experimental crossover corresponding to the appearance of the
Nernst signal. The calculated contour plot of the spinon pairing
parameter is compared with the Nernst signal contour
plot,\cite{ong2} showing a good agreement.

Finally, at an even lower temperature, the SC transition
temperature $T_c$, both holon pairs and RVB pairs, hence also the
hole pairs, become coherent.  It will be shown that the phase
coherence is established via a phase transition of a planar gauged
quantum XY-type, almost identical to that of the classical 3D
XY-model. The SC transition temperature is calculated as a
function of doping concentration, and is compared with the scaled
value of the resonance mode energy, observed in neutron
experiments\cite{Yu2009} to show the universal ratio between these
two quantities anticipated from our theoretical treatment.

The rest of the paper is organized as follows. Section II is a brief
introduction to our semionic spin-charge gauge approach, to make the
paper more self-contained. Section III is devoted to the holon pairing
mechanism. In section IV we discuss the spinon pairing, while in section
V the SC transition is considered.  Discussions and conclusions are
given in section VI. Several technical derivations are outlined in
Appendices.  A preliminary report of the present work has already
appeared in Ref.~\onlinecite{epl}.

\section{The spin-charge gauge approach}
\subsection{Slave semions}
 In this subsection we present an outline of our slave-particle approach
  using the ``semionic'' spin-charge decomposition, applicable only in 2D (and 1D) systems.
We assume that the main features of the low-energy physics of the
hole-doped cuprates can be captured by the $t-J$ model
\begin{eqnarray}
\label{eq:1}
\hat{H}_{t-J} &=&P_G[ -t\sum_{\langle i,j \rangle,\sigma}
\hat{c}^{\dagger}_{i,\sigma} \hat{c}_{j,\sigma}+h.c.-\mu \sum_j \hat{n}_{j} \nonumber\\
&&+ J \sum_{\left\langle
    i,j \right\rangle}  (\vec{S}_i \cdot \vec{S}_j+ \frac{1}{4}
\hat{n}_i \hat{n}_j)]P_G,
\end{eqnarray}
where $P_G$ is the Gutzwiller projection imposing
no-double-occupation condition and the lattice sites correspond to
those of the Cu atoms in the CuO$_2$ planes of the cuprates. The
particle number and spin operators are defined as
\begin{eqnarray}
\label{eq:2}
\hat{n}_i= \sum_{\sigma}\hat{c}^{\dagger}_{i\sigma} \hat{c}_{i\sigma},
\hspace{0.5cm} \vec{S}_i = \sum_{\alpha\beta}\hat{c}^{\dagger}_{i\alpha}
\frac{\vec{\sigma}_{\alpha\beta}}{2} \hat{c}_{i\beta}\;.
\end{eqnarray}
$t$, $J$, and $\mu$ in Eq.~\eqref{eq:1} are hopping amplitude,
spin exchange and chemical potential, respectively. The hole
operator carries both charge and spin degrees of freedom, with
no-double-occupation constraint. Formally they can be treated
separately by the standard slave-particle approach,
$\hat{c}_{i\sigma} = \hat{b}_{i\sigma}\hat{h}^{\dagger}_i$, where
$\hat{h}_i$ is a fermionic holon operator, and $\hat{b}_{i\sigma}$
is a bosonic spinon operator. The no-double occupation condition
is automatically ensured by the spinless fermion, while the
correct counting of degrees of freedom (dof) is imposed by the
constraint $\sum_{\sigma}
\hat{b}^{\dagger}_{i\sigma}\hat{b}_{i\sigma} = 1$ on the $\hat{b}$
field, so that
${n}_i=\hat{h}_i\hat{h}^{\dagger}_i=1-\hat{h}^{\dagger}_i\hat{h}_i.$
At half filling the charge degree of freedom is frozen, then the
slave-particle transformation reduces to the standard
Schwinger-boson approach.

In 2+1 dimensional systems, one can bind statistical fluxes to
particle-excitations, resulting \emph{only} in a change of the
statistics. This is achieved in the Hamiltonian formalism by
minimally coupling the matter fields to suitable composite
``statistical gauge field operators''. The introduction of these
fluxes in the lagrangian formalism is materialized via statistical
Chern-Simons gauge fields. In our case, holes carry both charge
and spin degrees of freedoms, so we associate two statistical
gauge fields  with hole operators, one of which is a U(1) gauge
field $\bf{B}$ coupled to the holon $\hat{h}$-field and related to
the charge, while the other  is an SU(2) field $\bf{V}$ coupled to
the spinon $\hat{b}$-fields and related to the spin. By carefully
choosing the coupling constants of the corresponding Chern-Simons
terms, we can keep the original hole field still fermionic. In the
Hamiltonian formalism, the statistical gauge fields operators can
be chosen as follows:
\begin{eqnarray}
\label{eq:3}
&&B_{\mu}(\vec{x}) = \frac{1}{2} \sum_{\vec{l}} \hat{n}_{\vec{l}}
\partial_{\mu} \arg(\vec{x}-\vec{l}), \nonumber\\
&&V^{a}_{\mu}(\vec{x})\sigma^a = \frac{1}{i} e^{-i \sum_l \hat{S}^{b}_l \sigma^b
  \arg(\vec{x}-\vec{l})}\partial_{\mu} e^{i \sum_l \hat{S}^{b}_l \sigma^b
  \arg(\vec{x}-\vec{l})},
\end{eqnarray}
where the sums are carried over lattice sites $\vec{l}$, the sum
over the spin indices $(a,b=x,y,z)$ is understood hereinafter and
the function $\arg(\vec{x})$ is the angle of the vector $\vec{x}$.
The corresponding U(1) and SU(2) fluxes, $\Phi_h$ and $\Phi_s$,
bound to the hole at site $\vec{j}$ are given by
\begin{eqnarray}
\label{fluxes}
&&e^{i\Phi_h(\vec{j})}=e^{i \int_{\vec{j}}^\infty d x^{\mu}
  B_{\mu}(\vec{x})}=e^{i \sum_{\vec{l}} \hat{n}_{\vec{l}}
  \arg(\vec{j}-\vec{l})},\nonumber\\
&&(e^{i\Phi_s(\vec{j})})_{\alpha\beta}=(P e^{i \int_{\vec{j}}^\infty d
  x^{\mu} V_{\mu}(\vec{x})} )_{\alpha\beta} \nonumber\\
&&\hspace{1.7cm}=(e^{i \sum_{\vec{l}} \hat{S}^b_{\vec{l}}
  \sigma^b \arg(\vec{j}-\vec{l})})_{\alpha\beta},
\end{eqnarray}
where $\alpha,\beta=1,2$ are the SU(2) spin indices. The
integration runs over a path joining $\vec{j}$ to infinity and $P$
denotes the path-ordering.  Binding the holon to the U(1)-flux
generated by $\bf{B}$ and the spinon to the SU(2)-flux generated
by $\bf{V}$ chosen as in Eq.~\eqref{eq:3} one obtains U(1) and
SU(2) invariant fields, respectively,  both obeying semionic
statistics,\cite{frohlich1992,marchetti1998} {\it i.e.} their
interchange produces a $\pm i$ factor, intermediate between the
bosonic $+1$ and the fermionic $-1$ case, whence the name
``semion''.\cite{girvin}

This ``semionic'' approach is quite suitable to study the physics
of holes dressed by a spin vortex as described in the Introduction
because the SU(2)-gauge field naturally incorporates the spin
vortices.  To show that  $V^{a}_{\mu}$ is indeed the gauge field
associated with spin vortices let us consider the simplest case of
one hole located at $\vec{l}$ with spin $\hat{S}^{a}$. Then
$V^{a}_{\mu}(\vec{x})$ simplifies to $\hat{S}^{a}\partial_{\mu}
\arg(\vec{x}-\vec{l})$ and using $\epsilon_{\mu\nu} \partial_\mu
\partial_\nu \arg (\vec{x}-\vec{y}) = \delta(\vec{x}-\vec{y})$ we
get
\begin{eqnarray}
\label{sv}
\epsilon^{\mu\nu}(\partial_\mu V^{a}_{\nu}(\vec{x}))=  \hat{S}^{a}
\delta(\vec{x}-\vec{l}).
\end{eqnarray}
Eq.~\eqref{sv} is a spin analogue of the  charged vortex
introduced by Laughlin in the fractional quantum Hall effect and,
in fact, a semionic representation of the electron was advocated
originally by him in the early days of high temperature SC
boom.\cite{lau}

\subsection{Improved Mean Field Approximation}
\label{sec:improved-mean-field}
In this subsection we sketch the key approximations involved in
our approach to the ``normal'' state; one of these approximations
(the optimization procedure) appears rather unconventional in
slave-particle approaches.

Being too difficult to be solved exactly, the gauge field approach
outlined above provides a reasonable base of an improved mean
field analysis\cite{marchetti1998} that, dimensionally reduced,
works quite well for one dimensional $t-J$
model,\cite{marchetti1996} reproducing correctly also the
non-trivial critical exponents of its correlation functions (the
spin-vortices become kink strings in 1D). In two dimensions, this
mean field theory involves an optimization of the spin
configuration around holons dressed by vortices, although it can
be carried out only approximately and not rigorously as in the
one- dimensional case. In the improved \emph{semionic} mean field
approximation (MFA), the spinon configurations around holons are
optimized leading to a new bosonic spinon on the optimized
spinon-background, denoted by $\hat{z}$, which is therefore
different from the $\hat{b}$-field in the standard slave fermion
approach, but still satisfying the constraint
$\hat{z}^{\dagger}_{i\alpha}\hat{z}_{i\alpha}=1$.  From now on it
is this spinon that we refer to. In the adopted MFA we neglect the
holon fluctuations in $\bf{B}$ and the spinon fluctuations in
$\bf{V}$. This leads to a much simpler form of the two statistical
gauge fields denoted by $\bar{\bf{B}}$ and $\bar{\bf{V}}$,
respectively, to distinguish them from the exact values. The
$\bar{\bf{B}}$ field is actually a static one, without dynamics,
\begin{eqnarray}
\label{eq:4}
\bar{B}_{\mu}\approx  \frac{1}{2}\sum_l
\partial_{\mu}\arg(\vec{x}-\vec{l}),
\end{eqnarray}
and it provides a $\pi$-flux phase factor $e^{i\bar{B}_{ij}}$ per
plaquette for the holon field because
$\int_{\square}\bar{B}_{ij}=\pi$.  For the SU(2) gauge field only
the $\sigma_z$-component survives:
\begin{eqnarray}
\label{eq:5}
\bar{V}^z_{\mu}(x) \approx -\sum_l
\hat{h}^{\dagger}_l\hat{h}_l\frac{(-1)^{|l|}}{2} \partial_{\mu}
\arg(\vec{x}-\vec{l}),
\end{eqnarray}
with a pure-gauge static term being gauged away.  Note that there
is the holon number operator  in the right hand side of
Eq.~\eqref{eq:5}, which means that the spin vortex is always
centered on the hole, and its topological charge (named chirality)
is $(-1)^{|l|}$ depending on the parity of the site index, where
$|l|=l_x+l_y$. The effect of the optimal spin flux is then to
attach a spin-vortex to the holon, with opposite chirality on the
two N\'eel sublattices, and the rigidity holding up a vortex being
provided by the AF background.  These vortices take into account
the long-range quantum distortion of the AF background caused by
the insertion of a dopant hole, as first discussed in
Ref.~\onlinecite{ss}. As in the one-dimensional case the
optimization involves also a spin-flip associated to every holon
jump between different N\'eel sublattices, hence in the $t$-term
the spinons appear in the ``ferromagnetic'' Affleck-Marston
(AM)\cite{am} form $\hat{\chi}^{s}_{ij} =(
\hat{z}^{\dagger}_ie^{iV^N_{ij}\sigma_z}\hat{z}_j)^{\#(i)}$, where
$\#(i)$ denotes complex conjugation if $i$ belongs to the ``odd''
sublattice, with a phase ambiguity left by the optimization,
whereas in the $J$-term it appears in the ``AF'' RVB form
$\hat{\Delta}^{s}_{ij}= \epsilon^{\alpha\beta} \hat{z}_{i\alpha}
(e^{iV^N_{ij}\sigma_z}\hat{z}_{j})_{\beta}$, where
\begin{eqnarray}
\label{eq:6}
V^N_{ij} = \int_i^jdx^{\mu} \bar{V}_{\mu}^z(\vec{x})\approx
\bar{V}^z_{\mu} (\frac{\vec{i}+\vec{j}}{2}).
\end{eqnarray}
The above AM/RVB dichotomy is peculiar to the semion approach
involving the SU(2) spin rotation group even in 1D, where it can
be rigorously derived. It does not appear in the standard U(1)
slave fermion or boson approaches.

In the above MFA the hole field operator can be decomposed as a
product of the holon and the spinon operators along with fluxes:
\begin{eqnarray}
\label{eq:7} \hat{c}_{i\sigma} = \hat{h}_i^{\dagger} e^{i
\Phi^h_i} (e^{i \Phi^s_i} \hat{z}_{i})_{\sigma}.
\end{eqnarray}
%This leads us to a spin-charge gauge approach
%whose scheme is outlined below, and for details one is referred to
%Refs.\onlinecite{marchetti1998}.  This method has also been applied to
%describe the transport and thermodynamic properties of cuprate
%superconductor in the pseudogap
%region\cite{marchetti2004a,marchetti2004b,marchetti2005,marchetti2008}.

The resulting MFA of the $t-J$ model Eq.~\eqref{eq:1} is written in
terms of holon fields $\hat{h}_i$ and spinon field $\hat{z}_i$ as
\begin{widetext}
\begin{eqnarray}
\label{eq:8} \hat{H}_{t-J} \approx t\sum_{\left\langle i,j
\right\rangle} \hat{h}^{\dagger}_j
e^{i\bar{B}_{ij}}\hat{h}_i\hat{\chi}^{s}_{ij} +h.c -\mu\sum_i
\hat{h}_i^{\dagger}\hat{h}_i + \frac{J}{2} \sum_{\langle i,j
\rangle} (1-\hat{h}^{\dagger}_i\hat{h}_{i}
-\hat{h}^{\dagger}_j\hat{h}_{j})
\hat{\Delta}_{ij}^{s\dagger}\hat{\Delta}^{s}_{ij}
+\hat{h}_i^{\dagger}\hat{h}_i \hat{h}^{\dagger}_j \hat{h}_j
\hat{\Delta}_{ij}^{s\dagger}\hat{\Delta}^{s}_{ij}.
\end{eqnarray}
\end{widetext}

The Euclidean Lagrangian used in the path-integral formalism is
then obtained by replacing the field operators
$\hat{h},\hat{h}^{\dagger}$ and $\hat{z},\hat{z}^{\dagger}$ with
Grassmann $(h,h^*)$ and complex number $(z,z^*)$, respectively,
and adding the time-derivative terms
\begin{eqnarray}
\label{time}
\sum_i h^*_i \partial_0 h_i +(1-h^*_i h_i)(-1)^{|i|}z^*_i \partial_0 z_i.
\end{eqnarray}

The Hamiltonian Eq.~\eqref{eq:8} is our starting point for
describing the High $T_c$ cuprate SC. At the mean field level, the
first two terms describe the motion of the holons, which are
coupled to the spinons through the AM factor whose modulus we
treat as a constant, giving a small correction to the hopping
amplitude $t$ of holons that we neglect. Its phase factor  $\sim
e^{i\theta_{ij}}$ instead cannot be neglected, and it provides a
gluing force between the spinon and holon. Then the mean field
Hamiltonian of holon reads
\begin{eqnarray}
\label{eq:10-1}
\hat{H}^{0}_h = t\sum_{\left\langle i,j \right\rangle}
\hat{h}^{\dagger}_j e^{i(\bar{B}_{ij}+\theta_{ij})}\hat{h}_i +h.c -\mu\sum_i
\hat{h}_i^{\dagger} \hat{h}_i.
\end{eqnarray}

In two-dimensional bipartite lattices for fermions in magnetic
field the optimal flux per plaquette is $\pi$ at half-filling
(Lieb theorem\cite{lieb}) and numerically it is also true for
close fillings at low temperatures, whereas it is zero
sufficiently far away from half-filling. Therefore the optimal
flux in a plaquette for $(\bar{B}_{ij}+\theta_{ij})$ is arguably
$\pi$ for small doping and low temperatures, and 0 for
sufficiently high dopings and/or high temperatures. We conjectured
that this corresponds to the crossover between the ``pseudogap
phase''(PG) and the ``strange metal phase''(SM) as varying the
doping or temperature in the cuprates, where PG is the ``lower
pseudogap'' in the literature identified with the inflection point
in resistivity and the broad peak in the specific heat coefficient
$\gamma$. This conjecture is supported by the comparison of the
behavior of the theoretically derived crossover temperature $T^*
  \approx \frac{1}{9 \pi} |\ln \delta|$,\cite{marchetti2007}
  with experimental data, where the appearance of
$|\ln \delta|$ is due to the long-range tail of spin vortex
interactions. Therefore we fix the phase ambiguity left by the
optimization in the AM term by choosing this phase zero for PG
since $\bar{B}$ has already $\pi$ flux and is opposite to
$\bar{B}$ for SM to effectively cancel $\bar{B}$.

If we replace the holon density by its average in MFA, the third
term in Eq.~\eqref{eq:8} describes the motion of $z$-spinons with
$J$ renormalized to $\tilde{J}\equiv J(1-2\delta)$. Without
doping, using the identity
\begin{eqnarray}
\label{ide}
|\hat{\Delta}^{s}_{ij}|^2 + |\hat{\chi}^{s}_{ij}|^2 =1,
\end{eqnarray}
holding for bosonic spinons, together with Eq.~\eqref{time} in the
continuum limit it leads to a standard nonlinear $\sigma$-model
describing the low energy physics of the AF background. With
doping the spinons are scattered by holons dressed by spin
vortices and that leads to a short range AF correlation. Such a
process is revealed by expanding the SU(2) phase factor inside the
RVB factor in the third term of Eq.~\eqref{eq:8} to the second
order, obtaining in the continuum limit, self-consistently in the
region with unbroken SU(2) spin symmetry
\begin{eqnarray}
\label{eq:9}
\tilde J \int d^2x \bar{V}^{z2}_{\mu}(\vec{x})
z^{*}_{\alpha}(x)z_{\alpha}(x).
\end{eqnarray}
In MFA we replace $\bar{V}^{z2}_{\mu}$, positive definite by definition,
by a statistical average.  The spatial average of
$\bar{V}^{z2}_{\mu}(\vec{x})$ at fixed holon position $\vec{x}_i$ by
using Eq.~\eqref{eq:5} reads
\begin{eqnarray}
\label{mass} \sum_{\vec{x_i},\vec{x_j}} (-1)^{|i|+|j|}
\triangle^{-1}(\vec{x}_i-\vec{x}_j),
\end{eqnarray}
where $|i|\equiv |\vec{x}_i|$ and $\triangle$ is the two
dimensional lattice Laplacian.\cite{marchetti1998}
Eq.~\eqref{mass} appears as the energy of a two dimensional
Coulomb gas with the lattice spacing as an ultraviolet cutoff,
which can be evaluated at fixed density $\delta$ by a quenched
approximation leading to a doping-dependent mass term for spinon,
which in the low doping limit is given by
\begin{eqnarray}
\label{ms}
m_s^2(\delta) =\langle \bar{V}^{z2}_{\mu}\rangle \approx  \frac{1}{2}
|\delta\ln\delta|,
\end{eqnarray}
consistent with AF correlation length ($\xi_{AF} \sim (m_s)^{-1}$)
at small $\delta$ extracted from the neutron experiments.\cite{ke}
In $\xi_{AF}$ the factor $\delta^{-\frac{1}{2}}$ is just the mean
distance between holes, while the factor $|\ln\delta|$  comes from
the long-range tail of the vortex interactions and it turns out to
be a key feature in many physical quantities in our approach.  The
spinon gap is also crucial for eliminating the over-counting of
low-energy degrees of freedom often encountered in slave-particle
approaches, giving rise to problems in the computation of
thermodynamic quantities.\cite{hl} In fact, because of the spinon
gap, the low-$T$ thermodynamics in our approach is essentially
dominated by the gapless holons, while the contributions of the
transverse and scalar gauge fluctuations to the free energy almost
cancel each other.\cite{marchetti2008}  In the Lagrangian form,
our massive $\sigma$-model derived from Eqs.~\eqref{time},
~\eqref{eq:9} and ~\eqref{ms} can be conveniently written as
\begin{eqnarray}
\label{eq:10}
\mathscr{L}_s &=&  \frac{1}{g} \int d^3x
[|(\partial_0-iA_0)z_{\alpha}|^2 -
v_s^2|(\partial_\mu-iA_\mu)z_{\alpha}|^2 \nonumber\\
&&+ m_s^2(\delta) z^{*}_{\alpha}z_{\alpha}](x),
\end{eqnarray}
where  an implicit momentum cutoff is implied inside the magnetic
Brillouin zone (MBZ), $g=8\tilde{J}a^2$ with $a$ lattice spacing,
$v_s=\tilde{J}a$, and the emergent gauge field $A_{\mu}$ is
generated by the fluctuations of spinons:
\begin{eqnarray}
\label{eq:11}
A_{\mu} \approx e^{i \vec{Q} \cdot \vec{x}}\frac{1}{i}
z_{\alpha}^{\dagger}(x)\partial_{\mu}z_{\alpha}(x) + ...,
\end{eqnarray}
with $\vec{Q}$ the AF wave vector, and it corresponds to the long
wavelength limit of $\theta_{ij}$, the phase factor of AM factor
$\chi_{ij}$. Note that in the massive $\sigma$-model
Eq.~\eqref{eq:10}, the constraint $z^{\dagger}z=1$ on the
$z$-field is relaxed. Holons and spinons are coupled by the gauge
field $A_{\mu}$, giving rise to overdamped resonances for holes
and magnons with strongly $T$-dependent
life-time.\cite{marchetti2007} This dependence originates from the
dynamics of the transverse mode of the gauge field that is
dominated by the contribution of the gapless holons. Their Fermi
surface generates an anomalous skin effect, with momentum scale
\begin{equation}
\label{reizer}
Q \approx (T k_F^2)^{1/3},
\end{equation}
known as the Reizer momentum,\cite{reizer1989a,reizer1989b} where
$k_F$ is the holon Fermi momentum measured from the Dirac point in
$\pi$-flux phase. For the appearance of Reizer skin effect the
presence of a gap for spinons is crucial, because gapless spinons
would Bose condense at low $T$ thus gapping the gauge field
through the Anderson-Higgs mechanism and destroying the
$T$-dependent skin effect that reduces the coherence of hole and
magnon.  The transport physics of PG is dominated by the interplay
between the short-range AF order due to spinons and the thermal
diffusion induced by the gauge fluctuations triggered by the
Reizer momentum, producing in turn the metal-insulator
crossover.\cite{marchetti2007}

More generally, the above semionic mean field treatment based upon
a spin-charge gauge approach to the $t-J$ model provides a
description of many transport and thermodynamic properties of High
Tc cuprates in PG
region\cite{marchetti1998,marchetti2004b,marchetti2004a,marchetti2005,marchetti2007,marchetti2008}
whose doping-temperature behavior is in qualitative, and sometimes
even semi-quantitative, agreement with experimental data. In the
following, we  present details of the novel non-BCS mechanism for
high Tc SC outlined in the Introduction. Here we just rewrite the
SC order parameter in the approximation adopted above:
\begin{eqnarray}
\label{eq:12}
\hat{\Delta}^c_{ij} = \epsilon_{\alpha\beta} \hat{c}_{i\alpha}
\hat{c}_{j\beta} \sim \hat{h}^{\dagger}_i \hat{h}^{\dagger}_j
\hat{\Delta}^s_{ij} e^{i\bar{B}_{ij}},
\end{eqnarray}
which can be obtained by Hubbard-Stratonovich transformation in
the path integral formalism.  In the next section we discuss the
holon pairing $\langle \hat{h}^{\dagger}_i
\hat{h}^{\dagger}_j\rangle$, while in section
\ref{sec:spinon-rvb-pairing} the spinon pairing
$\langle\hat{\Delta}^s_{ij} \rangle$.

\section{Holon Pairing}
\label{sec:holon-pairing}
\subsection{Holon Hamiltonian with Attractive Interaction}
\begin{figure}[htpb]
  \centering
  \includegraphics[width=5cm]{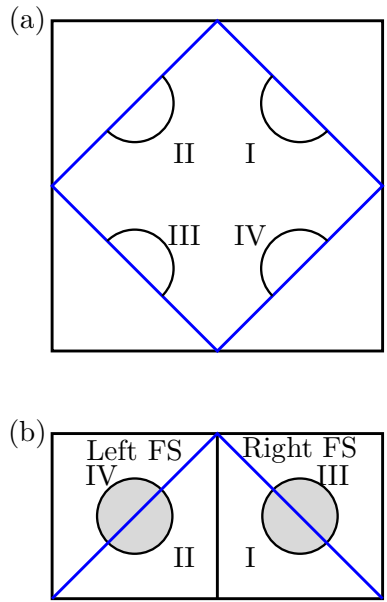}
  \caption{ The Brillouin zone and Fermi surface of free holon with
    $\pi$ flux. The folded MBZ with blue lines as boundary in (a) is
    equivalent to the rectangular in (b).}
  \label{fig:1}
\end{figure}

The Hamiltonian Eq.~\eqref{eq:10-1} in PG describes the motion of
holons which are subjected to a staggered $\pi$-flux field and the
gauge field $\theta_{ij}$,  coupling them to spinons. To get the
low energy physics of holon, we first neglect the gauge field
$\theta_{ij}$ generated by spinons, and it will be reinserted (in
an approximate form) by Peierls substitution.  The remaining terms
can be solved exactly. We find that the holon spectrum involves
two Dirac cones due to the presence of the $\pi$-flux (Hofstadter
mechanism).\cite{hof} The Fermi surface of holon is a small one
with Fermi wave-vector $k_F\approx\pi\delta$.\cite{marchetti2004b}
Due to the staggered $\pi$-flux, we divide the square lattice into
two sublattices, A(even sites) and B(odd sites). On each
sublattice, the holon's annihilation operators are denoted by
$\hat{a}$ and $\hat{b}$, respectively.
% To realize the $\pi$-flux per
% plaquette, we fix the gauge of $\bar{B}_{ij}$ in the following way
% \begin{eqnarray*}
% &&\hat{a}^{\dagger}_i \hat{b}_{i\pm\vec{e}_1} \rightarrow \hat{a}^{\dagger}_i
% \hat{b}_{i\pm\vec{e}_1} e^{i\pi/4} \nonumber\\
% &&\hat{a}^{\dagger}_i \hat{b}_{i\pm\vec{e}_2} \rightarrow \hat{a}^{\dagger}_i
% \hat{b}_{i\pm\vec{e}_2} e^{-i\pi/4},
% \end{eqnarray*}
% that is, when the holon jumps from B site to A site, it collects a
% phase factor $e^{i\pi/4}$ if jumping along $x$-direction, and
% $e^{-i\pi/4}$ if jumping along $y$-direction.
The Hamiltonian Eq.~\eqref{eq:10-1} of free holon can then be
recast in a quadratic form
\begin{eqnarray}
\label{eq:13} \hat{H}_{0}^h \sim \sum_{\vec{k}} (t_{\vec{k}}
\hat{a}^{\dagger}_{\vec{k}} \hat{b}_{\vec{k}} + h.c.) -
\mu\sum_{\vec{k}} (\hat{a}_{\vec{k}}^{\dagger} \hat{a}_{\vec{k}} +
\hat{b}_{\vec{k}}^{\dagger} \hat{b}_{\vec{k}} ),
\end{eqnarray}
where the momentum runs within the magnetic Brillouin zone (MBZ)
and $t_{\vec{k}} =2t(\cos k_x e^{i\pi/4} + \cos k_y e^{-i\pi/4})$.
It is straightforward to obtain the spectrum $\epsilon(\vec{k}) =
\pm |t_{\vec{k}}|$, with the Fermi surface consisting of four half
circles around $(\pm\pi/2,\pm\pi/2)$, as shown in
Fig.~\ref{fig:1}a, where the blue lines are the boundary of MBZ.
The Fermi energy is $t\delta$ approximately.  There are two
primitive reciprocal vectors, $\vec{\pi}_{\pm}\equiv (\pm\pi,\pi)$
by which we can translate the MBZ in the 3rd and 4th quadrants to
get another equivalent rectangular one as shown in
Fig.~\ref{fig:1}b, which consists of two Dirac cones centered
around $\vec{Q}_L=(-\pi/2,\pi/2)$ (left) and $\vec{Q}_R=
(\pi/2,\pi/2)$ (right), respectively. In this transformation, we
note that $ \hat{a}_{\vec{k}+\vec{\pi}_{\pm}} = \hat{a}_{\vec{k}}$
and $\hat{b}_{\vec{k}+\vec{\pi}_{\pm}} = -\hat{b}_{\vec{k}}$,
where a minus sign appears for $\hat{b}$-field defined on odd
sublattice,
 but the form of Hamiltonian Eq.~\eqref{eq:13} is still
  invariant, because $t_{\vec{k}}$ also changes sign after translation.

Accordingly, all the holon operators can be labeled by the
``flavor'' index $\alpha =L, R$ distinguishing left
  and right Dirac zones, so does the Hamiltonian of free holons
  $\hat{H}^h_{0}=\sum_{\alpha=L,R}\hat{H}^h_{0,\alpha}$ with
\begin{eqnarray}
\label{eq:14}
\hat{H}^h_{0,\alpha}&=&\sum_{\vec{k}} (t_{\alpha,\vec{k}}
\hat{a}^{\dagger}_{\alpha,\vec{k}} \hat{b}_{\alpha,\vec{k}}+h.c.) \nonumber\\
&&- \mu(\hat{a}_{\alpha,\vec{k}}^{\dagger}
\hat{a}_{\alpha,\vec{k}} +\hat{b}_{\alpha,\vec{k}}^{\dagger}
\hat{b}_{\alpha,\vec{k}}),
\end{eqnarray}
where $ t_{R,\vec{k}} \approx 2t(-k_x+ik_y)$ and $t_{L,\vec{k}} \approx
2t(k_x+ik_y)$. In Eq.~\eqref{eq:14}, the momentum $\vec{k}$ only takes
values in the range $[-\pi/2,\pi/2]\times [-\pi/2,\pi/2]$, which is one
quarter of the original BZ.

Now we consider the holon-holon interactions. As shown
  in Sec.~\ref{sec:improved-mean-field}, the last term in
  Eq.~\eqref{eq:8} is \emph{repulsive} for holons which  \emph{cannot}
  be the pairing force between holons, and is usually negligible in the low
  doping limit. Meanwhile, the third term in Eq.~\eqref{eq:8} implies an
  effective long-range interaction between the holons mediated by the
  spin vortices bound to holons, which turns out to be attractive
between different N\'eel sublattices. This leads to the
instability of holons towards pair formation and \emph{ is our key
attractive force.}
%  Physically it is due to
%the \textcolor{red}{quantum} distortion of the AF background
%caused by the holes.
Such an effect in the simplest form was first realized by S.
Trugman\cite{tr} in the early days of High Tc research, who
pointed out that putting two holes next to each other on two
N\'eel sublattices would save energy $J$.  We include this effect
in MFA by introducing a term coming from the average of $z^\dagger
z$ in ~\eqref{eq:9} obtaining:
\begin{equation}
\label{zh}
%\tilde{J}  \langle z^* z \rangle
J_{\text{eff}} \sum_{i,j} (-1)^{|i|+|j|} \triangle^{-1}(i -j)
\hat{h}^{\dagger}_i \hat{h}_i \hat{h}^{\dagger}_j \hat{h}_j.
\end{equation}
In the static approximation for holons, Eq.~\eqref{zh} describes a 2D
lattice Coulomb gas with charges $\pm 1$ depending on the N\'eel
sublattice and coupling constant $J_{\text{eff}}=\tilde J \langle
z^\dagger z \rangle$, where the average $\langle z^{\dagger}z\rangle$
can be estimated from the free spinon spectrum (which will be given in
the next section, see Eq.~\eqref{eq:47} by setting $\Delta_0^{s}=0$)
with the following result
\begin{eqnarray}
\label{eq:15}
J_{\text{eff}} &=& \tilde{J}\int d^2\vec{q}(q^2+m_s^2)^{-1/2} \nonumber\\
&=& J(1-2\delta)(\sqrt{\Lambda^2+m_s^2}-m_s),
\end{eqnarray}
where $\Lambda$ is the momentum cutoff for spinon excitations. For 2D
Coulomb gases with the above parameters a pairing appears for a
temperature $T_{ph} \approx {J_{\text{eff}}} /2\pi $ (a more precise
estimation is given later, in fact, $T_{ph}$ is the upper PG crossover
temperature determined by $\Delta_0^h(k_F)$ of Eq.~\eqref{eq:26}) which
turns out to be inside the SM ``phase''. Hence the whole PG ``phase''
lies below $T_{ph}$. However, we will discuss only the SC arising from
the PG phase, anticipating that extrapolation to SM phase will introduce
only quantitative changes (actually the role of a next nearest neighbor
hopping $t'$-term appears relevant in SM\cite{gam}).  In the large-scale
limit the two dimensional Coulomb interaction gives rise to a screening
effect, with a screening length $\ell_s$ which in the Thomas-Fermi
approximation is proportional to $1/\sqrt{\delta}$.\cite{coul} In view
of the above considerations we approximate the large-scale effective
potential in momentum space by
\begin{eqnarray}
\label{eq:16}
V_{\text{eff}}(\vec{p})
  \approx \frac{J_{\text{eff}}}{p^2+\ell_s^{-2}}.
\end{eqnarray}
The large-scale holon interaction then has the following simplified form
\begin{eqnarray}
\label{eq:17}
H^h_{I} &\sim&
%  - \sum_{i,j} V_{\text{eff}}(\vec{i}-\vec{j})
% \hat{a}^{\dagger}_i \hat{b}^{\dagger}_j \hat{b}_j \hat{a}_i \nonumber\\
% &=&
- \sum_{\vec{p}_{1}\vec{q}_{1}\vec{p}_{2}\vec{q}_{2}}
V_{\text{eff}}(\vec{q}_{1} -\vec{q}_{2})\nonumber\\
&&\times\delta(\vec{p}_{1}-\vec{p}_{2}+\vec{q}_{1}-\vec{q}_{2})
\hat{a}^{\dagger}_{\vec{p}_{1}} \hat{b}^{\dagger}_{\vec{q}_{1}}
\hat{b}_{\vec{q}_{2}} \hat{a}_{\vec{p}_{2}}.
\end{eqnarray}
% and in BCS approximation it becomes
% \begin{eqnarray}
% \label{BCS} H^h_{I} &\sim& - \sum_{\vec{p},\vec{q}}
% V_{\text{eff}}(\vec{p}-\vec{q}) \hat{a}^{\dagger}_{\vec{p}}
% \hat{b}^{\dagger}_{-\vec{p}} \hat{b}_{-\vec{q}} \hat{a}_{\vec{q}}.
% \end{eqnarray}

Due to the long range tail of vortex-vortex interaction, the pairing
strength for large momentum($\vec{q}\sim(\pi,0)$) transfer between
different Dirac cones is much smaller than that for small
momentum($\vec{q}\sim 0$) transfer. Hence, in the presence of
interaction, the left and right flavors of holons can still be decoupled
approximately. Considering the BCS approximation, where pairing occurs
between holons in states with opposite momentum, one obtains the
decoupled Hamiltonians $H^h_{I,\alpha}$ for each flavor $\alpha$,
\begin{eqnarray}
\label{eq:18}
\hat{H}^h_{I,\alpha}=- \sum_{\vec{p},\vec{q}} V_{\text{eff}}(\vec{p}-\vec{q})
\hat{a}^{\dagger}_{\alpha,\vec{p}} \hat{b}^{\dagger}_{\alpha,-\vec{p}}
\hat{b}_{\alpha,-\vec{q}} \hat{a}_{\alpha,\vec{q}}.
\end{eqnarray}
%We just show that the holons can be approximately described by two flavors (L and
%R) of Dirac fermions decoupled completely with each other, within
%each of which there is an attractive interaction. As known, the low energy
%excitation of high $T_c$ cuprate superconductor are experimentally found
%inside the area enclosed by the red lines as in Fig.~\ref{fig:1}a,
%around the Fermi arcs. In our formalism, the four regions separated by
%red lines are reunited to be two small Fermi surface as in
%Fig.~\ref{fig:1}b.

We shall now focus only on the quasiparticles near the Fermi circles,
which allows us to make the following gauge transformations for the
holon operators with different flavors separately
\begin{eqnarray}
\label{eq:19} \hat{a}_{\alpha,\vec{k}} \rightarrow
\hat{a}_{\alpha,\vec{k}}e^{i\theta_{\alpha,\vec{k}}/2},
\hspace{0.2cm} \hat{b}_{\alpha,\vec{k}} \rightarrow
\hat{b}_{\alpha,\vec{k}}e^{-i\theta_{\alpha,\vec{k}}/2},
\end{eqnarray}
where the angles $\theta_{\alpha,\vec{k}}$ are chosen to cancel the
phase of $t_{\vec{k}}$ so that the kinetic term reads
\begin{eqnarray}
\label{eq:20}
\hat{H}^h_{0,\alpha} \approx v_F k(\hat{a}^{\dagger}_{\alpha,\vec{k}}
\hat{b}_{\alpha,\vec{k}} + h.c.)
\end{eqnarray}
with $v_F = 2t$.
%Note that the pairing term transforms like
%\begin{eqnarray}
%\label{eq:21}
%\hat{b}_{\alpha,-\vec{q}} \hat{a}_{\alpha,\vec{q}} \rightarrow
%\hat{b}_{\alpha,-\vec{q}} \hat{a}_{\alpha,\vec{q}} e^{i\pi/2}
%\end{eqnarray}
%therefore, the interaction term Eq.~\eqref{eq:53} is gauge invariant.
Eqs.~\eqref{eq:18} and ~\eqref{eq:20} are our basic equations to
describe the pairing of holons.

\subsection{D-wave Pairing}
In this subsection, we show that the \emph{$d$-wave} pairing symmetry is
composed naturally of two $p$-wave pairing within the left and right
Dirac cones, an idea first proposed by Sushkov {\it et
  al.}\cite{sus1,sus2} in a different setting.  The corresponding
pairing parameter has a form respecting the $C_{4v}$ rotation symmetry,
\begin{eqnarray}
\label{eq:22}
\Delta^h_{\alpha,\vec{k}} = \left\{
  \begin{array}{ll}
    \Delta^h(k) \frac{k_x-k_y}{k}, \text{ if }\alpha=R, \\
    \Delta^h(k) \frac{-k_x-k_y}{k}, \text{ if }\alpha=L,
  \end{array}
\right.
\end{eqnarray}
where the momentum $\vec{k}$ is measured from $\vec{Q}_{R,L}$,
respectively, and the magnitude of the order parameter is the same for
both $R$ and $L$ flavors. Note that we are now working with the
rectangular magnetic Brillouin zone (see Fig.\ref{fig:1}b) and the
$p$-wave pairing takes place within the two circular Fermi surfaces. If
transformed back to diamond magnetic Brillouin zone as in
Fig.~\ref{fig:1}a, the order parameters in region III and IV change
their signs due to the fact that $\hat{b}_{\vec{k}+\vec{\pi}_{\pm}} =
-\hat{b}_{\vec{k}}$, which leads to a perfect d-wave pairing in the
original Brillouin zone.

Applying the standard BCS treatment we get the following MF Hamiltonian
\begin{eqnarray}
\label{eq:23} \hat{H}^h_{\alpha} = \hat{H}^h_{0,\alpha} +
\sum_{\vec{k}} (\Delta^{h}_{\alpha,\vec{k}}
\hat{a}^{\dagger}_{\alpha,\vec{k}} \hat{b}^{\dagger}_{\alpha,
-\vec{k}} + h.c.),
\end{eqnarray}
where the order parameter satisfies the gap equations,
\begin{eqnarray}
\label{eq:24}
&&\Delta^h_{\alpha,\vec{k}} = \sum_{\vec{q}} V_{\text{eff}}(\vec{k}-\vec{q})
\frac{\Delta_{\alpha,\vec{q}}^h}{2\epsilon_{\alpha,\vec{q}}} \tanh
\left(\frac{\epsilon_{\alpha,\vec{q}}}{2T} \right).
\end{eqnarray}
It turns out that Eq.~\eqref{eq:23} has two decoupled branches of
solutions (see Appendix \ref{sec:diag-mf-hamilt} for details). One of
them with higher energy without FS provides a matrix element suppressing
the spectral weight of the original holon field $\hat{h}$ outside the
MBZ as in PG.\cite{marchetti2004b} The other one is responsible for the
low energy physics of holon pairing which we will focus on in the
following and its spectrum has a simple BCS form
\begin{eqnarray}
\label{eq:25}
\epsilon^h_{\vec{k}} = \sqrt{(v_Fk-\mu)^2 + |\Delta_{\vec{k}}|^2}.
\end{eqnarray}
As common for non-weakly coupled attractive Fermi systems, the MF
temperature at which $\Delta^h$ becomes non-vanishing should be
identified with the pairing temperature $T_{ph}$.

For brevity, we consider the $p$-wave order parameter in the right
cone, which has the form $\Delta^{h}_{\vec{k}} =
\Delta_0^h(k)(\cos\theta_{\vec{k}}-\sin\theta_{\vec{k}})$.  The
radial part $\Delta_0^h(k)$ is decoupled from its angular part
approximately (see Appendix \ref{sec:diag-mf-hamilt}), which is
plotted in Fig.~\ref{fig:deltah} for different values of the
screening length $\ell_s$. We observe that  holons near the Fermi
surface take part in pairing which results in a peak of
$\Delta^h(k)$ centered around $k\sim k_F$. Actually, the number of
holons participating in pairing is determined by the screening
length $\ell_s$. If we increase $\ell_s$, a higher percentage of
holons can interact with each others at longer distance, the peak
of $\Delta^h(k)$ in Fig.~\ref{fig:deltah} becomes higher and
wider, that implies a bigger fraction of holons is involved in
pairing. A more rigorous treatment would actually involve taking
into account self-consistently the UV cutoff and chemical
potential change, as discussed {\it e.g.} in Refs.
\onlinecite{schmitt1989,randeria1990}, but for simplicity we
refrain to do that, assuming that our system is sufficiently
BCS-like and our treatment catches already the key behavior, as
Fig.~\ref{fig:deltah} suggests.
% Notice however that decreasing $\delta$
%the density of holons, and hence $k_F$, decreases.
%It is even possible that all the holons can
%participate in pairing as long as the screening of the vortex
%interaction is weak enough.

\begin{figure}[htbp]
%\centerline{\includegraphics[width=8.5cm]{delta.ps}}
\centerline{\includegraphics[width=8.5cm]{figure2a.eps}}
\centerline{\includegraphics[width=8.5cm]{figure2b.eps}}
\caption[]{\label{fig:deltah} Plots of pairing gap of holon as functions
  of momentum in the upper panel, and as functions of temperature in the lower
  panel for different screening length $\ell_{s}$. $\Delta_0^h$ is
  plotted in units of $t$. It is seen that the holons near the Fermi
  surface take part in pairing, leading to a peak of $\Delta(k)$ center
  around $k\sim k_F$. }
\end{figure}

The maximum value of the order parameter at zero temperature can be
taken as the typical energy scale of pairing strength, which is the
value of $\Delta_0^h$ at the Fermi momentum. Though it is difficult to
get an analytical solution of $\Delta_0^h(k_F)$ from the radial gap
equation (see Eq.~\eqref{eq:74}), one can get an approximate expression
for it as a function of the parameters $J_{\text{eff}}$, $k_F$ and
$\ell_s$, which has the following form
\begin{eqnarray}
\label{eq:26} \Delta^h_0(k_F) \approx 0.06 J_{\text{eff}}
(k_F\ell_s) \exp \left( -
  \frac{40\mu}{J_{\text{eff}} (k_F\ell_s)^{2}} \right),
\end{eqnarray}
being not sensitive to the energy cutoff as long as the screening
length $\ell_s$ is larger than $1/\Lambda$.

Now we can write down the $d$-wave order parameters near the
original four Fermi arcs (see Fig.~\ref{fig:1}a)
\begin{itemize}
\item $\Delta^h_{\vec{k}}\approx v_{\Delta} (k_x-k_y)/\sqrt{2}$ in
  quadrant I,
\item $\Delta^h_{\vec{k}}\approx v_{\Delta} (-k_x-k_y)/\sqrt{2}$ in quadrant II,
\item $\Delta^h_{\vec{k}}\approx v_{\Delta} (-k_x+k_y)/\sqrt{2}$ in quadrant III,
\item $\Delta^h_{\vec{k}}\approx v_{\Delta} (k_x+k_y)/\sqrt{2}$ in quadrant IV,
\end{itemize}
where $v_{\Delta}\equiv \sqrt{2}\Delta^h_0(k_F)/k_F$.

So far we discussed the $d$-wave paring symmetry in the momentum
space in the long wavelength limit, and now we check that when
extrapolating the result to the lattice scale we recover the
desired pairing symmetry in real space. Computing the nearest
neighbor pairing between site $\vec{x}$ and $\vec{x}+\vec{\delta}$
we get,
\begin{eqnarray}
\label{eq:27} \langle \hat{b}_{\vec{x}}
\hat{a}_{\vec{x}+\vec{\delta}}\rangle \approx\frac{1}{V}
\sum_{\vec{k},\alpha} [\langle \hat{b}_{\alpha,-\vec{q}}
\hat{a}_{\alpha,\vec{q}} \rangle e^{i\vec{Q}_{\alpha}\cdot
\vec{\delta}}] e^{i\vec{k}\cdot \vec{\delta}},
\end{eqnarray}
where $V$ is the volume of the system and the summation over
$\vec{k}$ is in the range $[-\pi/2,\pi/2]\times [-\pi/2,\pi/2]$.
Note that $\langle \hat{b}_{\alpha,-\vec{q}}
\hat{a}_{\alpha,\vec{q}} \rangle$ has the same symmetry as
$\Delta^h_{\alpha,\vec{q}}$ (see Eq.~\eqref{eq:69}), then by using
Eq.~\eqref{eq:22}, one can easily prove $\langle \hat{b}_{\vec{x}}
\hat{a}_{\vec{x}+\vec{\delta}}\rangle =\langle \hat{b}_{\vec{x}}
\hat{a}_{\vec{x}-\vec{\delta}}\rangle$ and $\langle
\hat{b}_{\vec{x}} \hat{a}_{\vec{x}+\vec{e}_1}\rangle=-\langle
\hat{b}_{\vec{x}} \hat{a}_{\vec{x}+\vec{e}_2}\rangle$,
% $\Delta^h_{\vec{\delta}}=\Delta^h_{-\vec{\delta}}$, and
% $\Delta^h_{\vec{e}_1} = -\Delta^h_{\vec{e}_2}$,
which are the typical
features of d-wave order parameters in real space.

\subsection{Nodal approximation and Gauge Invariance}

In the BCS approximation discussed in the previous subsection the
holon is gapless only at the 4 nodal points of
$\Delta^h_{\vec{k}}$. However, in a large-scale gauge-invariant
treatment whereas one can keep the modulus of the order parameter
$\Delta^h$ as in BCS, we must include its spatially dependent
phase, which we denote by $\phi^h(x)$. (A precise procedure to go
from the lattice to the continuum phase field is discussed in
Ref.~\onlinecite{tesanovich2008}.) The effects of $\phi^h(x)$ on
holons is non-trivial and will be discussed in detail in
Ref.~\onlinecite{gam}. However, to derive our basic RVB gap
equation in the next section it turns out that we can assume
consistently that $\phi^h(x)$ doesn't break the nodal structure.
In fact the nodal structure appears if we neglect the phase
fluctuations, so that the holon pairs are assumed condensed.
According to Refs.
\onlinecite{nozieres1985,botelho2006,tempere2009} this is the
correct procedure to deal with the gap equation for the modulus of
the order parameter. However, if holon pairs are only formed but
not yet condensed, it is incorrect to identify $\Delta^h$ as the
gap for holons (see Ref.~\onlinecite{gam}).
\begin{figure}[htbp]
%\centerline{\includegraphics[width=6cm]{figure2.ps}}
\centerline{\includegraphics[width=5cm]{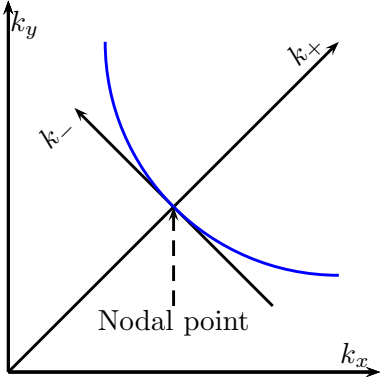}}
\caption[]{\label{fig:2} The coordinate system $(k_{+},k_{-})$ taken
  in the nodal approximation. This is for the first quadrant.}
\end{figure}
In this subsection we utilize the Peierls substitution to take the
gauge fields back into account around the nodal points, in
agreement with the above remarks. In the nodal approximation the
momenta are expanded around the nodes in the four quadrants of the
MBZ.  In Fig.~\ref{fig:2}, we plot the nodal coordinate system in
the 1st quadrant, where
\begin{eqnarray}
\label{eq:28}
k_x = \frac{k_{+}+k_{-}}{\sqrt{2}}, \hspace{0.2cm} k_y =
\frac{k_{+}-k_{-}}{\sqrt{2}} .
\end{eqnarray}
In terms of $k_{+}$ and $k_{-}$, using the the gap dependence on
momentum $\vec{k}$ obtained in the last subsection, the energy
spectrum around the node of the 1st quadrant is simply $\pm
\sqrt{v_F^2 k_{+}^2 +
  v_{\Delta}^2 k^2_{-}}$, which arises from the nodal Hamiltonian in the
1st quadrant
\begin{eqnarray}
\label{eq:29}
\mathcal{H}^h_{\text{1st nodal}} = v_F k_{+} \sigma_z +
v_{\Delta}k_{-}\sigma_y,
\end{eqnarray}
which reproduces the spectrum of the gapless nodal
excitation. Therefore, adding also the contribution of the phase
$\phi^h(x)$ of the order parameter, a large-scale h/s gauge invariant
Hamiltonian in real space reads
\begin{eqnarray}
\label{eq:30}
\mathcal{H}^h_{\text{1st}} = \begin{pmatrix}
 -i\partial_{+} -A_{+}+A_0 & -e^{i\phi^h}\partial_{-}
  \\
e^{-i\phi^h}\partial_{-}  & i\partial_{+} -A_{+}-A_0
\end{pmatrix},
\end{eqnarray}
where the emergent gauge field $A_{+}$ and $A_0$(see Eq.~\eqref{eq:11})
is reinserted, and the parameters $v_F$ and $v_{\Delta}$ are omitted for
the sake of simplicity. There is an obvious U(1) redundancy of this
Hamiltonian. Let us denote the nodal Dirac quasi-particles field by
$\chi_\alpha(x)$. The $h/s$ gauge transformation $ \chi \rightarrow \chi
e^{i\Lambda},\phi^h\rightarrow \phi^h+2 \Lambda, A_{\mu} \rightarrow
A_{\mu}+\partial_{\mu}\Lambda$
% \begin{eqnarray}
% \label{eq:31}
% \chi \rightarrow \chi e^{i\Lambda},\phi^h\rightarrow \phi^h+2 \Lambda, A_{\mu}
% \rightarrow A_{\mu}+\partial_{\mu}\Lambda
% \end{eqnarray}
leaves Eq.~\eqref{eq:30} invariant provided that $\Lambda$ is time
independent.  We then make the field redefinition from $\chi_{\alpha}$
to $\tilde{\chi}_{\alpha}$ as $\tilde{\chi}_1=\chi_1 e^{i\phi^h/2}$ and
$\tilde{\chi}_2= \chi_2 e^{-i\phi^h/2}$,
% \begin{eqnarray}
% \label{eq:32}
% \tilde{\chi}_1=\chi_1  e^{i\phi^h/2}, \hspace{0.2cm}
% \tilde{\chi}_2= \chi_2  e^{-i\phi^h/2}
% \end{eqnarray}
so that the nodal field becomes neutral under h/s gauge
transformations, a `nodon'.\cite{sf}
%\textcolor{red}{PUT A REFERENCE? M.Fisher et
 % al. Phys. Rev. B 60 (1999) 1654}.\textcolor{red}{I DON'T KNOW if PUT
 % THIS IN A NOTE We refrain to discuss the $\bf{Z}_2$ ambiguity involved
  %here and in the redefinition of the spinon field before
 % ~eqref{eq:56}[\textbf{By ye: possibly being Eq.~\eqref{eq:50}}], analogous to that in
%?PUT? \cite{sf}[T. Senthil and Matthew P. A. Fisher, Phys. Rev. B 63,
 %134521 (2001)], since it appears irrevelevant in the following.}
%the unitary transformation
%\begin{eqnarray*}
%\mathcal{H}_{\text{1st}}\rightarrow \begin{pmatrix}
% e^{-i\phi^h/2} & 0 \\ 0 & e^{i\phi^h/2}
%\end{pmatrix} [\mathcal{H}_{\text{1st}} + i \partial_0] \begin{pmatrix}
 %e^{i\phi^h/2}& 0 \\ 0 & e^{-i\phi^h/2}
%\end{pmatrix}
%\end{eqnarray*}
The above redefinition leads to a more convenient form of the nodal
Hamiltonian:
\begin{eqnarray}
\label{eq:33}
\mathcal{H}_{\text{1st}}&=&\begin{pmatrix}
 -i\partial_{+}-a_{+}+a_0 & -\partial_{-}\\
\partial_{-}  & i\partial_{+} -a_{+}-a_0
\end{pmatrix} \nonumber\\
&=&-a_{+}+(-i\partial_{+}+a_0)\sigma_3-i\partial_{-}\sigma_2,
\end{eqnarray}
where the h/s gauge invariant field $a_{\mu}=A_{\mu}-\frac{1}{2}\partial_{\mu}
\phi^h$ is introduced.
%It is obvious that the Hamiltonian is h/s gauge invariant, which
%is necessary since h/s gauge field is actually artificial which should
%not enter into the physical quantities.

Rotating the coordinate by $\pi/2$ successively, one may get the
nodal Hamiltonian in the other three quadrants.

\subsection{Effective Action of $a_{\mu}$ }
In this subsection, we turn to the path-integral formalism and
derive an effective action (needed to discuss RVB gap equation)
for $a_{\mu}$ in the nodal approximation by integrating out the
holon fields.  In the 1st quadrant, the effective Lagrangian in
the Euclidean space for nodal quasi-particles is given by:
\begin{eqnarray}
\label{eq:34} &&\mathscr{L}^{\text{1st}} =
\bar{\chi}(x)[\gamma^{\mu}(\partial_{\mu}-ib^{\text{1st}}_{\mu})
]\chi(x),
\end{eqnarray}
where $\gamma^{\mu}=\{\sigma_x,-\sigma_{y},\sigma_z\}$, $\partial_{\mu}=
\{\partial_0,\partial_{+},\partial_{-} \}$ and
$b^{\text{1st}}_{\mu}=\{-ia_{+},ia_0, 0 \}$.
% \begin{eqnarray}
% \label{eq:35}
% &&\gamma^{\mu}=\{\sigma_x,-\sigma_{y},\sigma_z\}, \nonumber\\
% &&\partial_{\mu}= \{\partial_0,\partial_{+},\partial_{-} \}, \nonumber\\
% &&b^{\text{1st}}_{\mu}=\{-ia_{+},ia_0, 0 \}.
% \end{eqnarray}
%Obviously, $\{\gamma^{\mu},\gamma^{\nu}\} =
%2\delta^{\mu\nu}\mathbf{I}$.
The effective action for $b^{\mu}$ (at $T=0$) is defined as
\begin{eqnarray}
\label{eq:36}
&&S^{1st}_{\text{eff}}[a^{\mu}] = -
\ln\det[\gamma^{\mu}(\partial_{\mu}-ib^{\text{1st}}_{\mu}) ] \approx\nonumber\\
&& -\frac{1}{2} \int d^2k \int d\omega
[b^{\text{1st}}_{\mu} \Pi^{\text{1st}}_{\mu\nu}b^{\text{1st}}_{\nu}](\vec{k},\omega).
\end{eqnarray}
By adapting the calculations of Ref. \onlinecite{sharapov}, the
leading terms of the bubbles for small $\omega,
|\vec{k}|,\omega/|\vec{k}|$  behave like
\begin{eqnarray}
\label{eq:37} \Pi^{\text{1st}}_{00}\sim c_1|\vec{k}|,
\hspace{0.2cm} \Pi^{\text{1st}}_{++}\sim c_2, \hspace{0.2cm}
\Pi^{\text{1st}}_{0+}\sim 0.
\end{eqnarray}

%(at non-vanishing temperature $\Pi^{\text{1st}}_{00}$ acquire a term linear in $T$).
The effective action in the other three quadrants is similar to that in
Eqs.~\eqref{eq:36} and \eqref{eq:37}. For example, the 3rd quadrant can
be obtained by rotating the coordinate by $\pi$, therefore by changing
$a_{\pm}\rightarrow -a_{\pm}$ and
$\partial_{\pm}\rightarrow-\partial_{\pm} $, we can obtain the
corresponding bubble $\Pi_{\mu\nu}$ and gauge field $b_{\mu}$. Note that
the coordinate transformation does not involve the time axis. Then we
have
\begin{itemize}
\item $\Pi_{\mu\mu}^{\text{3rd}} = \Pi^{\text{1st}}_{\mu\mu}$, if
  $\mu=0,1,2$,
%\item $\Pi_{ij}^{\text{3rd}} = \Pi^{\text{1st}}_{ij}$, if $i,j=1,2$,
%\item $\Pi_{0i}^{\text{3rd}} =- \Pi^{\text{1st}}_{0i}$, if $i=1,2$,
\item
  $b^{\text{3rd}}_{\mu}=\{ia_{+},ia_0,0\}.$
\end{itemize}
Similar procedure can be applied to the 2nd and 4th quadrants.

After summing over all four quadrants, in the quadratic
approximation, we have the effective action
%\begin{widetext}
\begin{eqnarray}
\label{eq:38}
{S}^h_{\text{eff}}[a_{\mu}]&=&\int d^{2}k\int  d\omega  \sum_{i=+,-} \nonumber\\
&&[a_i\Pi_{00}a_i
+a_{0}\Pi_{ii}a_{0}-2a_{0}\Pi_{0i}a_{i}](\vec{k},\omega).
\end{eqnarray}
%\end{widetext} .
Using Eq.~\eqref{eq:37} and its analogues one sees that
Eq.~\eqref{eq:38} is a variant of the effective action for
QED$_3$.

%%%%%%%%%%%%%%%%%%%%%%%%%%%%%%%%%%%%%%%%%%%%%%%%%%%%%%%%%%%%%

\section{Spinon RVB Pairing}
\label{sec:spinon-rvb-pairing}
\subsection{Mean Field Lagrangian with Spinon Pairing}
   In this subsection we derive the mean field Lagrangian for
spinon pairing in the presence of holon pairs.

In the PG region, the spinon part in the $t-J$ model can be
described by the massive sigma model in CP$^1$ form
Eq.~\eqref{eq:10}, and the four fermion interaction term $\sim J
(\hat{h}^{\dagger}_i\hat{h}^{\dagger}_j
\hat{\Delta}^{s\dagger}_{ij})(\hat{h}_j\hat{h}_i\hat{\Delta}^s_{ij})
$ (see the last term in Eq.~\eqref{eq:8}), is simply neglected for
small doping $\delta$ in considering the normal state, because it
is proportional to $\delta^2$. Note this interaction term is
positive (for $J>0$), hence repulsive due to the \emph{semionic}
mean field approach, contrary to the usual fermionic case.
%This is
%also consistent within our whole scheme.
%Because the spinon is described
%by a bosonic $\sigma$-model without the constraint $|z|^2=1$, if
%$J<0$(attractive), we are led to a catastrophe as
%$|\Delta^s_{ij}|\rightarrow\infty$.
However, once the holon pairing is stabilized, the gauge interaction
between holon and spinon, overcoming the above repulsion forces the
spinons to form singlet-RVB pairs and the above term becomes relevant.
To investigate the spinon pairing, one can apply a Hubbard-Stratonovich
transformation to the four fermion interaction term, obtaining
\begin{eqnarray}
\label{eq:39} \sum_{\langle ij\rangle}
-\frac{2|\Delta^s_{ij}|^2}{J \tau^2} + \Delta^{s*}_{ij}
\epsilon^{\alpha\beta} z_{i\alpha}z_{j\beta} + h.c.,
\end{eqnarray}
where $\tau\equiv|\langle\hat{h}_i \hat{h}_j\rangle|$ and in MFA
\begin{eqnarray}
\label{eq:40}
\Delta^s_{ij} = \frac{J}{2}\tau^2 \langle \epsilon^{\alpha\beta} \hat{z}_{i\alpha}
\hat{z}_{j\beta}\rangle= \frac{J}{2}\tau^2 \langle \hat{\Delta}^s_{ij}\rangle.
\end{eqnarray}
%which is assumed to be s-wave like. Correspondingly, a d-wave like
%pairing amplitude for holes reads
%\begin{eqnarray}
%\label{eq:41}
%\Delta^c_{ij} = \Delta^s_{ij}/\langle \hat{h}_i\hat{h}_{j}\rangle.
%\end{eqnarray}
%Here one may worry about the validity of Eq.~\eqref{eq:41} since
%$\langle \hat{h}^{\dagger}_i \hat{h}^{\dagger}_{j}\rangle$ may be zero
%leading to singularity.
%Notice that the spinon pairing must vanish if  $\langle \hat{h}^{\dagger}_i \hat{h}^{\dagger}_{j}\rangle=0$
%in this
%case too, so does $\Delta^c_{ij}$, otherwise the h/s gauge invariance is
%broken which is not allowed under any circumstances.
 In the continuum limit we get the Lagrangian for spinon
with a singlet spinon pairing
\begin{eqnarray}
\label{eq:42}
\mathscr{L}_{s} &=&
\sum_{\mu=0,1,2}z^{*}_{\alpha}[(\partial_{\mu}-iA_{\mu})^{2}+m^2_s]z_{\alpha}
\nonumber\\
&&+ \sum_{i=1,2}\Delta^{s*}_{i}(\vec{x})
\epsilon^{\alpha\beta} z_{\alpha}(\vec{x})\partial_{i}
z_{\beta}(\vec{x}) + h.c.,
\end{eqnarray}
where the index $i$ in $\Delta^{s*}$ labels the spatial directions and
we set $g$ and $v_s$ to 1 for convenience. (The spatial derivative term
in the square brackets has an implicit `-' sign, see Eq.~\eqref{eq:10}.)
As for the holon case, one can rewrite approximately the spinon pairing
as $\Delta_i^s(\vec{x})=\Delta_{i,0}e^{i\phi^s(\vec{x})}$ where $\phi^s$
is the phase of the spinon pairing amplitude. The Lagrangian
Eq.~\eqref{eq:42} is invariant under the h/s gauge transformation
$z_{\alpha}\rightarrow z_{\alpha}e^{i\Lambda}$, $A_{\mu} \rightarrow
A_{\mu} + \partial_{\mu}\Lambda$ and $\phi^s \rightarrow \phi^s + 2
\Lambda$. It is not convenient to deal with the off-diagonal terms in
the Lagrangian $\mathcal{L}_s$, hence we transform the spinon field from
$z_{\alpha}$ to $\tilde{z}_{\alpha}$ as $\tilde{z}_1=z_1
e^{i\phi^s/2}$,$ \tilde{z}_2= z_2 ^{*} e^{-i\phi^s/2}$ so that the
spinon field becomes neutral under h/s gauge transformations.  In terms
of the new fields $\tilde{z}_{\alpha}$, the spinon Lagrangian can be
written in a diagonal form $\mathscr{L}_s (x)= \tilde{z}^{\dagger}(x)
\Gamma_s(x) \tilde{z}(x)$
% \begin{eqnarray}
% \label{eq:43}
% \mathscr{L}_s (x)= \tilde{z}^{\dagger}(x) \Gamma_s(x) \tilde{z}(x)
% \end{eqnarray}
where the $2\times 2$ kernel $\Gamma_s$ reads (with $\Delta^s_{0,0}=0$)
\begin{widetext}
\begin{eqnarray}
\label{eq:44}
\Gamma_s = \sum_{\mu=0,1,2}-[\partial_{\mu}
-i(a_{\mu}+\frac{1}{2}\partial_{\mu}\phi)\sigma_z -
i\text{Im}(\Delta^s_{\mu,0})\sigma_x -i
\text{Re}(\Delta^s_{\mu,0})\sigma_y]^2 + m^2_s -|\Delta^s_{\mu,0}|^2,
\end{eqnarray}
\end{widetext}
with $\phi=\phi^h-\phi^s$ and
$a_{\mu}=A_{\mu}-\frac{1}{2}\partial_{\mu}\phi^h$, both of which are h/s
gauge invariant. The gradient of the $\phi$ field actually describes the
potential of standard magnetic vortices, since from Eqs.~\eqref{eq:12}
and ~\eqref{eq:40} $\phi$ is the phase of the condensate of hole-pairs.
% Note that we assume
%$\Delta^s_{0,0}=0$ if $\mu=0$, i.e., no pairing occurs in the temporal
%direction.

\begin{figure}[htbp]
\centerline{\includegraphics[width=6cm]{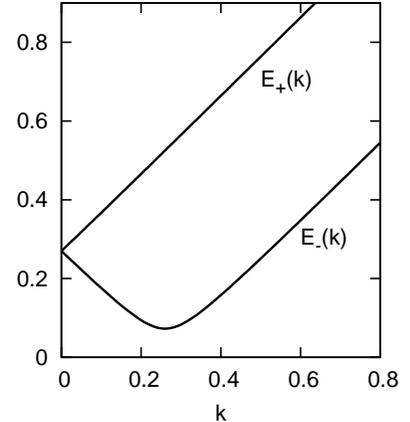}}
\caption[]{\label{fig:3} The spinon spectrum for $\delta=0.1$.}
\end{figure}

By neglecting  the gauge fields, one can work out the spinon
spectrum, which can be obtained from the zeros of the determinant
of the kernel $\Gamma_s$ in the momentum space:
\begin{eqnarray}
\label{eq:45}
%\det[\mathscr{D}^{-1}_s(\omega,k)]&=&
(-\omega^2+k^2+m^2_s)^2
%\nonumber\\
%&&
-4\sum_{i,j=1,2}\Delta^s_{i,0}\Delta^{s*}_{j,0}
k_i k_j =0.
\end{eqnarray}
We assume the rotational invariance for the spinon spectrum, which
requires
\begin{eqnarray}
\label{eq:46}
\Delta^s_{i,0}\Delta^{s*}_{j,0} + \Delta^{s*}_{i,0}\Delta^{s}_{j,0} =
2\delta_{ij} | \Delta^s_{i,0}|^2.
\end{eqnarray}
We can take $\Delta^s_{1,0} = \Delta_0^s$ and $\Delta^s_{2,0} =
\pm i\Delta_0^s$, where $\Delta^s_0$ can be {\it a priori} any
complex number, and both plus and minus signs are allowed for
$\Delta^s_{2,0}$.  Looking at the hole-pair order parameter
Eq.~\eqref{eq:12}, we see, however, that for consistency we have
to choose the constant phases of $\Delta^s_0$ equal to $-\bar
B_{ij}$ to ensure the correct symmetry, being the hole-pair
$d$-wave.
%Since the physical
%results do not depend on the phase of $\Delta_0^s$ and the sign of
%$\Delta^s_{2,0}$, for later convenience $\Delta_0^s$ is chosen to be
%positive number, and the plus sign is taken for $\Delta^s_{2,0}$.
 From Eq.~\eqref{eq:45} we
obtain the spectrum for spinon: it has
two (positive) branches
\begin{eqnarray}
\label{eq:47}
E_{\pm}(\vec{k}) = \sqrt{\vec{k}^2+m^2_s\pm 2\Delta_0^s |\vec{k}|}
\end{eqnarray}
which are plotted in Fig.~\ref{fig:3}.
% Note that a value of
%$\Delta^s_0$ larger than $m_s$ leads to the quasiparticles of
%$E_{-}(k)$ branch missing in the range of momentum $[\Delta^s_0-
%\sqrt{(\Delta^{s})^2 -m_s^2}, \Delta^s_0+ \sqrt{(\Delta^{s})^2
 % -m_s^2}]$, furthermore, the spin excitation of $E_{-}(k)$ branch is
%gapless.
%We think this is unphysical, and restrict $\Delta^s_0$ in the
%range $[0,m_s]$.
The positive branches of the dispersion are similar to those found
in a plasma of relativistic fermions,\cite{wel} which suggests the
following interpretation: if $|\Delta^s|\neq 0 $ the spinon system
contains a gas of RVB spinon pairs, an analogue of Coulomb neutral
pairs in the relativistic plasma, either in the plasma phase if
$\langle \Delta^s\rangle=0$, or in a condensate if $\langle
\Delta^s\rangle \neq 0$.  For a finite density of spinon pairs
there are two (positive energy) excitations, with different
energies, but the same spin and momenta. They are given, {\it
e.g.}, by creating a spinon up and by destructing a spinon down in
one of the RVB pairs. Notice that the minimum at $\tilde{J} |\vec
k|= |\Delta^s|$ in the lower branch is similar to the roton
minimum in superfluid helium and has an energy lower than $m_s$;
it implies a backflow of the gas of spinon-pairs dressing the
``bare spinon''. Hence RVB condensation would lower the spinon
kinetic energy. However, to make it occur one needs the gauge
contribution to overcome the spinon repulsion generated by the
Heisenberg term.

\subsection{Effective action of Gauge fields}
In this subsection, we derive the low-energy effective action of
$a_{\mu}$ and $\phi$ by integrating over the spinon fields.  For
this purpose we introduce a fictitious SU(2) gauge fields $Y_\mu$
as follows:
\begin{eqnarray}
\label{eq:48}
Y_\mu=\sum_{a=x,y,z} Y^a_{\mu} \frac{\sigma_a}{2}
\end{eqnarray}
with
\begin{eqnarray}
\label{eq:49}
Y^a_{\mu} = 2 \begin{pmatrix}
 0 & 0 & a_0+\partial_0\phi \\
\text{Im}(\Delta^s_{1,0}) & \text{Re}(\Delta^s_{1,0})&
a_1+\frac{1}{2}\partial_1\phi \\
\text{Im}(\Delta^s_{2,0}) & \text{Re}(\Delta^s_{2,0})&
a_2+\frac{1}{2}\partial_2\phi
\end{pmatrix}.
\end{eqnarray}
Then the kernel $\Gamma_s$ (see
Eq.~\eqref{eq:44}) can be written in a compact form
\begin{eqnarray}
\label{eq:50} \Gamma_s = \sum_{\mu}(\partial_{\mu} - i Y^a_{\mu}
\frac{\sigma^a}{2})^2 + M^2,
\end{eqnarray}
where we introduce the notation
\begin{eqnarray}
\label{M}
 M=\sqrt{m^2_s -2|\Delta^s_0|^2}
\end{eqnarray}
for convenience.
 After integrating over the spinon fields $z_{\alpha}$, one
obtains the effective action for $a_{\mu}$ and $\partial_{\mu}\phi$,
\begin{eqnarray}
\label{eq:51}
&&S_{\text{eff}}^s[\partial_{\mu}\phi,a_{\mu},\Delta^s_0] =
\ln\det(\Gamma_s) - \frac{2|\Delta_0^s|^2}{J\tau^2},
\end{eqnarray}
where the constant term comes from the Hubbard-Stratonovich
transformation. Since Eq. ~\eqref{eq:50} is formally describing a
relativistic 2-component boson of mass $M$ minimally coupled to the
SU(2) gauge field $Y^a_{\mu}$, the leading gauge invariant term is the
Yang-Mills Lagrangian, {\it i.e.}, the traced square ($\sum_{\mu\nu}
Y_{\mu\nu}^a Y_{\mu\nu}^a$) of the field strength $Y_{\mu\nu}$
\begin{eqnarray}
\label{eq:52}
Y_{\mu\nu} =
 \frac{\sigma^c}{2}[\partial_{\mu} Y_{\nu}^c - \partial_{\nu}
Y^c_{\mu} + \epsilon^{abc} Y^a_{\mu}Y^b_{\nu}]
\end{eqnarray}
and one easily computes, with $i=1,2$:
\begin{eqnarray}
\label{eq:53}
&&Y_{0i}^x=-(a_0+\partial_0\phi)\mbox{Re}(\Delta_{i0}^{s}),\nonumber\\
&&Y_{12}^x=\mbox{Re}(\Delta^s_{10})(a_2+\partial_2\phi)-\mbox{Re}(\Delta^s_{20})
 (a_1+\partial_1\phi),\nonumber\\
&&Y_{0i}^y=(a_0+\partial_0\phi)\mbox{Im}(\Delta_{i0}^{s}),\nonumber\\
&&Y_{12}^y=\mbox{Im}(\Delta^s_{20})(a_1+\partial_1\phi)-\mbox{Im}(\Delta^s_{10})
(a_2+\partial_2\phi),
\nonumber\\
&& y_{0i}^z=\partial_0(a_i+\partial_i\phi)-\partial_i(a_0+\partial_0\phi), \nonumber\\
&&y_{12}^z=\partial_1(a_2+\partial_2\phi)-\partial_2(a_1+\partial_1\phi),\nonumber\\
&&+\mbox{Im}(\Delta^s_{10})
\mbox{Re}(\Delta^s_{20}) - \mbox{Im}(\Delta^s_{20})\mbox{Re}(\Delta^s_{10})\nonumber.
\end{eqnarray}
Besides the Yang-Mills action there are also gauge non-invariant terms
which arise from the ultra-violet divergences of the continuous model
and must be included since the $x,y$ components of $Y^a_{\mu}$ are
actually constant.  For the 0th and 2nd order terms in $a_\mu$ and
$\phi$ we finally get
\begin{eqnarray}
\label{eq:54} &&S^{s,0}_{\text{eff}} =
-\frac{2|\Delta_0^s|^2}{{J}\tau^2} +\sum_{\omega,\vec{k}}
\ln[(\omega^2+E^2_{-}(\vec{k}))(\omega^2+E^2_{+}(\vec{k}))],
\nonumber\\
&& S^{s,2}_{\text{eff}} = \frac{1}{6\pi
  M}\{[\partial_{\mu}a_{\nu}-\partial_{\nu}a_{\mu}]^2 \nonumber\\
&&+|\Delta^s_0|^2
[2(a_0+\frac{1}{2}\partial_0\phi)^2+(\vec{a}+\frac{1}{2}\vec{\nabla}\phi)^2]\},
\end{eqnarray}
where a surface term ($\sim \partial_{1}a_{2}-\partial_{2}a_{1}$) has
been discarded.
%The coefficients are fixed by comparing with the exact integration by
%setting all gauge fields zero.
For $|\Delta^s_0|\neq 0$, $S^{s,2}_{\text{eff}} $ is the action of a
gauged XY or Stueckelberg model and the term in the last square bracket
is the celebrated Anderson-Higgs mass term.

\subsection{Gap equation of spinon pairing}
\label{sec:gap-equation-spinon} The gap equation is determined by
the saddle point of
$S_{\text{eff}}^s[a,\Delta_0^s]=S^{s,0}_{\text{eff}}[\Delta_0^s]+
S^{s,2}_{\text{eff}}[a,\Delta_0^s]+S^h_{\text{eff}}[a]$ with
respect to $\Delta_0^s$.  Note that since the interaction between
spinons is repulsive, it is crucial to take the gauge fluctuation
$S^{s,2}_{\text{eff}}$ into account, unlike in the traditional BCS
theory, where the electron interaction is attractive. To establish
the gap equation for the modulus of the order parameter, we
assume, as discussed {\it e.g.} in Refs.
\onlinecite{nozieres1985,botelho2006,tempere2009} for fermions,
that one should neglect the phase ($\phi$) fluctuations. Let us
also neglect for simplicity at first the holon contribution
$S^h_{\text{eff}}$, then the resulting gauge partition function,
denoted by $Z_{g}$,  is given by:
\begin{eqnarray}
\label{eq:55}
Z_{g} &=&\int D[a_{\mu}] e^{ -\int d^3x  \mathscr{L}_g[a_{\mu}]
}, \nonumber\\
\mathscr{L}_g&=&\frac{1}{3\pi
M}\left[a_{\mu}(-\partial^2g^{\mu\nu}
+\partial^{\mu}\partial^{\nu}+|\Delta_0^s|^2\lambda^{\mu\nu})a_{\nu}\right],
\end{eqnarray}
where $g^{\mu\nu}=\text{diag}(1,1,1)$ and
$\lambda^{\mu\nu}=\text{diag}(1,1/2,1/2)$ are $3\times3$ diagonal
matrices and a cutoff $\Lambda$  in both momenta and energy is
understood.
% Clearly, the mass term of the gauge fields is anisotropic in space time.
 The equation of motion of gauge field $a_{\mu}$ reads
\begin{eqnarray}
\label{eq:56}
-\partial^2a_{\mu}+\partial^{\mu}(\partial_{\nu}a_{\nu})+\lambda^{\mu\nu}a_{\nu}
= 0,
\end{eqnarray}
which implies that (without source) $a_{\mu}$ satisfies the equation
\begin{eqnarray}
\label{eq:57}
\lambda^{\mu\nu}\partial_{\mu}a_{\nu} = 0.
\end{eqnarray}
Eq.~\eqref{eq:57} is a constraint, meaning that the massive vector
bosons in two dimensions have two (physical) polarization modes.
Therefore, the calculation of the partition function of the vector
boson is not as trivial as integrating over $a_{\mu}$ directly, in
fact one must take care to count only the physical degrees of
freedom. The details of the evaluation of $Z_g$ is given in
Appendix \ref{sec:eval-part-funct}.
%following Ref.~\onlinecite{bazeia1983}
Here we only give the final result
\begin{eqnarray}
\label{eq:58}
Z_g &=&  \prod_{\omega,\vec{k}} \left[(3\pi M)^{\frac{3}{2}}
  \left( \omega^2+\frac{|\Delta_0^s|^2}{2} + \frac{\vec{k}^2}{2}
  \right)^{-\frac{1}{2}}\right. \nonumber\\
&&\hspace{1.0cm}\left.\times\left( \omega^2+\frac{|\Delta_0^s|^2}{2} +
    {\vec{k}^2} \right)^{-\frac{1}{2}} \right].
\end{eqnarray}
The first factor in r.h.s. of Eq.~\eqref{eq:58} contributes a
constant term to the free energy which is neglected in the
standard cases. However, in the present case it contains the
spinon order parameter $\Delta_0^s$, which can affect the total
energy as $\Delta_0^s$ varies, hence should be kept. Actually,
$3\pi M$ is the renormalization factor of the amplitude of the
gauge action, which can be absorbed into $a_{\mu}$ by rescaling
$a_{\mu}\rightarrow \sqrt{3\pi M}a_{\mu}$ with a Jacobian left for
the measure $D[a_{\mu}]$ in the energy-momentum space
\begin{eqnarray}
\label{eq:59} D[a_{\mu}]\rightarrow \prod_{\omega,\vec{k}}(3\pi
M)^{\frac{3}{2}}D[a_{\mu}],
\end{eqnarray}
where the power index $3/2$ is due to the fact that $a_{\mu}$ is a
three-dimensional vector.

In fact in the spinon gap equation the term $3\pi M$ already
balances the repulsive interaction. The contributions of the
spectrum of gauge quasiparticles, {\it i.e.}, the second and third
terms in Eq.~\eqref{eq:58}, do not change the spinon gap equation
qualitatively. Therefore, for simplicity we focus only on the
$M$-term, and the free energy including the contribution from the
h/s gauge fluctuation reads
\begin{eqnarray}
\label{eq:60}
&&\frac{1}{V}F_g[\Delta_0^s] \nonumber\\
&\approx& \frac{1}{\beta V} \sum_{\omega,\vec{k}}
\ln[(\omega^2+E^2_{-}(\vec{k}))(\omega^2+E^2_{+}(\vec{k}))] \nonumber\\
&& - \frac{3\Lambda^3}{4}\left[ \ln
m_s^2-\frac{2|\Delta_0^s|^2}{m_s^2} \right] - \Lambda^2
\frac{|\Delta_0^s|^2}{J \tau^2}.
\end{eqnarray}
%where we take the cutoff $\Lambda$ to be isotropic in energy-momentum space.
It is straightforward to obtain the gap equation by taking
derivative of $F_g$ with respect to $|\Delta_0^s|^2$:
\begin{eqnarray}
\label{eq:61}
0 &=&   \frac{3\Lambda^3}{2m^2_s}  -
\frac{\Lambda^2}{J\tau^2}\nonumber\\
&&-\frac{1}{2|\Delta_0^s| V} \sum_{\vec{k}}  \left[
\frac{k}{E_{-}\tanh \frac{E_{-}}{2T}}  - \frac{k}{E_{+}\tanh
\frac{E_{+}}{2T}}  \right].
\end{eqnarray}
The first term originates from the gauge action due to the
lowering of the spinon mass ($m_s \rightarrow
(m_s^2-|\Delta^s|^2)^{1/2}$), while the second term comes from the
original repulsive Heisenberg term and the last two terms are due
to the spinon excitations.  The first term in r.h.s. of
Eq.~\eqref{eq:61} is crucial, without which the gap equation has
no solution, since the last term is negative.  In
Eq.~\eqref{eq:61} only the value of $\tau$ is unknown, {\it i.e.},
the nearest neighbor holon pairing strength, which is a very short
range correlation and may not be accurate if being calculated via
the long wavelength pairing $\Delta_{\vec{k}}^h$ in momentum
space. However, we have already seen that extrapolating
$\Delta_{\vec{k}}^h$ to lattice scale, one gets the correct
symmetry in real space. Hence
%it should be
%proportional to $\Delta^h_0(k_F)$(see Eq.~\eqref{eq:26}), we then take
we take $\Delta^h_0(k_F)$ as the value of $\tau$ up to a scale
factor. Let us briefly comment on the relation of our RVB gap
equation Eq.~\eqref{eq:61} with that of the slave boson approach.
Whereas in the slave-boson approach the RVB pairs are made of
fermions and the Heisenberg term is attractive, so the
pair-formation is BCS-like, in our approach the RVB pairs are made
of bosons, and the Heisenberg term is repulsive, so the pair
formation arises from the decrease in the free energy of spinons,
via the lowering of their mass gap, induced by holon-pairing
through the gauge field.

% Notice that in our approach the leading part of the original
% Heisenberg term is used to provide the AF action for the spinons,
% using the identity ~\eqref{ide}(holding for bosonic spinons) .  Only
% the subleading term proportional to the holon-pair density is used
% to obtain the formation of a finite density of RVB-pairs in
% (\ref{gapeq}), so the derived SC can be view as vaguely
% reminiscent of Laughlin's Gossamer SC \cite{lau}.

So far we have not considered the vector boson quasiparticles,
whose spectrum has two branches as derived from Eq.~\eqref{eq:58},
\begin{eqnarray}
\label{eq:62}
&&E^{(1)}_{g}(\vec{k}) = \sqrt{k^2 + \frac{|\Delta_0^s|^2}{2}},
\hspace{0.2cm}E^{(2)}_{g}(\vec{k}) = \sqrt{\frac{k^2}{2} + \frac{|\Delta_0^s|^2}{2}}
\end{eqnarray}
and contributes to the gap equation with the following term
\begin{eqnarray}
\label{eq:63}
\sim \frac{1}{2 V}\sum_{\vec{k},n=1,2} \frac{1}{E^{(n)}_g\tanh \frac{E^{(n)}_g}{2T}}.
\end{eqnarray}
This contribution is positive and in balancing the gap equation
Eq.~\eqref{eq:61} plays a role similar to the $M$-term, that turns out
to be dominant.  It is interesting to note that Eq.~\eqref{eq:63} is
well defined in the gapped region $\Delta_0^s\ne 0$, and if
$\Delta_0^s=0$, it is proportional $\sim T\ln L$ which is divergent
unless $T=0$ if the typical length of the sample $L$ goes to
infinity. Such an infared divergence seems to imply a first order phase
transition when spinons begin to pair. However, this is {\it not the
  case}. In fact, when we take into account the contribution of holons
$S^h_{\text{eff}}$ to the action of gauge field (see Eq.~\eqref{eq:38}),
the dispersions ~\eqref{eq:61} become (see Appendix
\ref{sec:eval-part-funct}):
\begin{eqnarray}
\label{eq:64}
&&E^{(1)}_{g}(\vec{k}) = \sqrt{k^2 + f(\vec{k})+\frac{|\Delta_0^s|^2}{2}},\nonumber\\
&&E^{(2)}_{g}(\vec{k}) = \sqrt{\left( \frac{|\Delta_0^s|^2}{2}+f(\vec{k})
  \right) \left( 1+ \frac{\vec{k}^2}{|\Delta_0^s|^2+ \tilde{c}_2}
  \right)},
\end{eqnarray}
where $\tilde{c}_{2}=3\pi M c_{2}$ and $f(\vec{k})=3\pi M c_1
\sqrt{v_F^2k_{+}^2+v_{\Delta}^2k_{-}^2}$, and the divergence disappears.
% In the following discussion (but not in the numerical calculations) we
% keep the $M$-contribution in view of its
% dominance %and simplicity, neglecting the contribution from the gauge quasiparticles.

%\subsection{Phase diagram}
%In previous section, we established a gap equation Eq.~\eqref{eq:61}
% for spinon pairing by taking the gauge fluctuation into account.
In the low doping limit at $T=0$, expanding the last terms in the
r.h.s. of Eq.~\eqref{eq:61} we get
\begin{eqnarray}
\label{eq:65}
|\Delta_0^s|\approx \frac{\Lambda^{3/2}}{m_s^{1/2}}  \sqrt{1-
  \frac{m_s^2}{J\tau^2}}.
\end{eqnarray}
As the doping $\delta$ is decreased, $\tau$ goes to zero faster
than $m_s$, because the spinon mass $m_s^2\sim |\delta\ln\delta|$
and $\tau^2 \sim \delta e^{-\text{const.}}$(see
Eq.~\eqref{eq:26}), which implies that $|\Delta_0^s|$ has no
nonzero solution for sufficiently small doping. In other words,
there is a critical doping $\delta_{c}$ at zero temperature, below
which spinon pairing $\Delta_0^s$ must vanish. As the
non-vanishing of $\Delta_0^s$ is a pre-condition for SC, this
implies a critical doping for SC at $T=0$. On the other hand, at
the qualitative level, due to the cancellation of $\delta$ between
$m_s^2$ and $\tau^2$, if $\tau$ ( {\it i.e.} the holon-pairs
density) is sufficiently large Eq.~\eqref{eq:65} does have a
solution, because the remaining $|\ln\delta|$ is a decreasing
function. Notice again the crucial role of this logarithm, coming
from the long-range tail of spin-vortices.

\begin{figure}[htbp]
\centerline{\includegraphics[width=5.5cm]{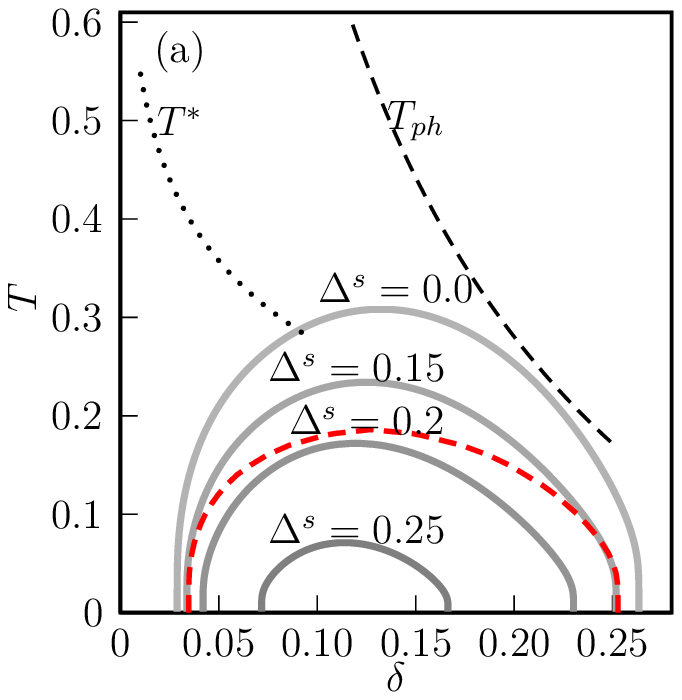}}
\centerline{\includegraphics[width=7.0cm]{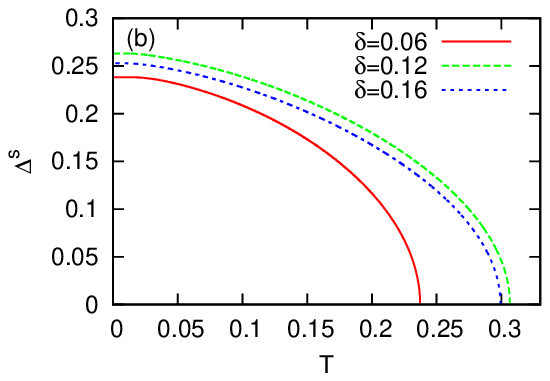}}
\caption[]{\label{fig:phasediagram}(Color online) (a) is the $T-\delta$
  phase diagram of the mean field gap equation of spinon for different
  values of MF spinon pairing $\Delta^s$ (gray lines) which could be
  compared with different levels of the Nernst signal\cite{ong1,ong2};
  $\Delta^s=0$ is $T_{ps}$.  (The curves at high dopings are not
  quantitatively reliable as they do not take into account the crossover
  to the ``strange metal'').  The dashed line is $T_{ph}$, the ``upper
  PG crossover temperature''.  The dotted line is the crossover
  temperature between the pseudogap and strange metal phases, $T^*$.
  (b) is the $\Delta^s$ as a function of temperature for fixed dopings.
  The temperature and $\Delta^s$ are in units of $J$. }
\end{figure}

At finite temperatures, we need to solve Eq.~\eqref{eq:61}
numerically. The crossover temperature at which in mean field
approximation $\Delta_0^s$ becomes non-vanishing is denoted by
$T_{ps}$ (not yet the SC $T_c$) and is related to the formation of
a finite density of RVB spinon pairs. From Eq.~\eqref{eq:26} we
see that to have solution for the gap equation we need
$\tau=\langle h_i h_j \rangle \sim \Delta_0^h \neq 0$,
consistently with the physical mechanism proposed, hence $T_{ph} >
T_{ps}$ and when the spinon RVB pairs are formed  together with
the already formed holon pairs, producing a finite density  of
preformed \emph{hole} pairs. Due to the $\phi$ phase fluctuations,
however, although the modulus of the SC order parameter $\Delta^c
\sim \Delta^s/\Delta^h$ of ~\eqref{eq:12} is non-vanishing, if the
hole pairs are not condensed one cannot interpret it as the hole
gap. The temperature dependence of $\Delta_s$ is presented in Fig.
~\ref{fig:phasediagram}b. One can see that, although near $T_{ps}$
the behavior is the typical square root of mean-field, at low $T$
it is definitely not BCS-like, never approaching a constant.

\section{Superconductivity}
Now we are ready to finally discuss the true SC transition.
\subsection{Nernst crossover}
In this subsection we first consider the physical effects due to a
finite density of hole pairs before their condensation.

The gauged XY or Stueckelberg model of Eq.~\eqref{eq:54} is well
known to have in the lattice two phases (see
Ref.~\onlinecite{seiler} for a rigorous discussion, while
Ref.~\onlinecite{kleinert} for a numerical analysis): Coulomb and
Higgs. If the coefficient, $\sim |\Delta_0^s|^2$ of the
Anderson-Higgs mass term for $a$ is sufficiently small, the phase
field $\phi$ fluctuates so strongly that it does not produce a
mass gap for $a_\mu$ and $\langle e^{i \phi}\rangle=0$ in the
Coulomb gauge (a gauge-fixing is necessary due to the Elitzur
theorem\cite{elitzur}).  This is the Coulomb phase, where a plasma
of magnetic vortices-antivortices appears. In the presence of a
temperature gradient a perpendicular external magnetic field
induces an unbalance between vortices and antivortices, giving
rise to a Nernst signal, even if the hole-pairs are not condensed
yet. Therefore we conjecture that this phase of the model
corresponds to the region in the phase diagram of underdoped
cuprates characterized by a non-SC Nernst signal and a comparison
between the experimental phase diagram in Refs.
\onlinecite{ong1,ong2} and the one derived in our model, supports
this idea.  The result is shown in Fig.~\ref{fig:phasediagram},
where the thick lines are equal-$\Delta_0^s$ lines.  One expects
that the level of $\Delta_0^s$ is roughly proportional to the
intensity of the Nernst signal and a comparison of the figure with
the experimental data\cite{ong1,ong2} shows a qualitative
agreement for the $\delta-T$ dependence. Note that the Nernst data
are strongly supported by the measured magnetic-field induced
diamagnetic signal,\cite{diam} as well as by STM visualized pair
formation\cite{yazdani} and quasi-particle
fingerprint.\cite{davis} The $T_{ph}$ line in the figure is the
upper pseudo-gap crossover temperature determined by
$\Delta_0^h(k_F)$ of Eq.~\eqref{eq:26}, hence it does not take
into account the transition to the SM phase, therefore can only be
taken as a qualitative trend.  At extremely low doping ($\delta
\lesssim 0.03$) the lines are not reliable because the quenched
approximation for vortices used in our approach is not valid for
too low vortex density.

\begin{figure}[htbp]
\centerline{\includegraphics[width=7cm]{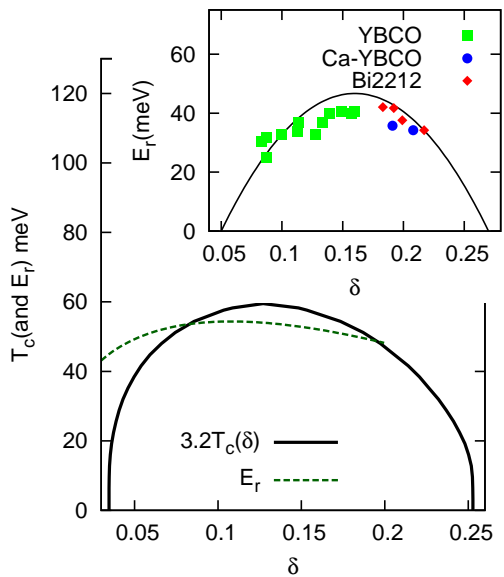}}
\caption[]{\label{fig:resonance}(Color online) The energy of the
  magnetic resonance $E_r$ estimated by $2m_s J$ for different dopings,
  compared with the scaled critical temperature $3.2T_c$. To compare
  with experiments we take $J=100meV$. The inset is the experimental
  results taken from Ref.~\onlinecite{Yu2009}, where the black solid
  line is a parabolic approximation to $T_c$ rescaled by 5.8.}
\end{figure}

\subsection{The superconducting transition}
Now we consider the true SC transition. For a sufficiently large
coefficient $|\Delta^s|^2$, the gauged XY or Stueckelberg model of
Eq.~\eqref{eq:54} is in the broken symmetry phase: the
fluctuations of $\phi$ are exponentially suppressed and $\langle e^{i
  \phi}\rangle \neq 0$ at $T=0$ or there is a quasi-condensation
(power-law-decaying order parameter) at $T>0$; accordingly
magnetic vortex-antivortex pairs become small and dilute, so the
gauge field is gapped.  At the same time the holon, and hence the
hole, acquires the nodal gap, {\it i.e.} the gap outside  the
nodes.  In fact, one can prove that, due to the fluctuations of
the field $\phi^h$, in our approach a gapless gauge field is
inconsistent with the coherence of holon pairs in PG, {\it i.e.,}
coherent holon pairs \emph{cannot
  coexist} with incoherent spinon pairs, as sketched in Appendix C. On
the other hand, due to the QED-like structure of holons-gauge action,
the gauge field cannot be gapped (in all components) by condensation of
holon pairs alone as shown by Eq.~\eqref{eq:37}; only the condensation
of RVB spinon pairs at the same time can open a gap to the gauge
fluctuations and then the nodal \emph{hole} gap. Thus as soon as $e^{i
  \phi^h}$ (quasi-)condenses, the same occurs to $\langle
h_ih_j\rangle$, so that SC emerges, since from Eqs.~\eqref{eq:40} and
~\eqref{eq:54} the SC order parameter is $\Delta^c\sim
\Delta^s/\langle h_ih_j\rangle \sim (\Delta^s_0/\Delta^h_0) e^{i \phi}$
and now its modulus and the expectation of its phase are nonzero (at $T
=0$, or power-law decaying at $T>0$). It follows that $T_c<T_{ps}$.

According to the above considerations, if we assume that the holon
contribution to the gauge field is subdominant, as expected, the SC
transition from the PG phase should occur roughly at a value of
$\Delta^s_0$ determined by the gauged XY model. Then one can extract an
estimate of the critical value of $\Delta^s_0$ from a formula presented
in Ref.~\onlinecite{kleinert} for the critical value of the coefficient
of Anderson-Higgs mass term in Eq.~\eqref{eq:54}. If one rescales the
gauge field $a_\mu$ to have the standard coefficient 1/2 for the Maxwell
term, denote by $q$ the charge of the $\phi$ field w.r.t. the rescaled
$a_\mu$ and denote by $\beta$ the coefficient of the Anderson-Higgs mass
term, then such formula reads:
\begin{eqnarray}
\label{klei}
\beta_c \approx (3-\frac{q^2}{4})^{-1},
\end{eqnarray}
where $\beta_c$ denotes the critical value. The value $q=0$ corresponds
to a pure XY model; in our case (see Eq.~\eqref{eq:54}) $q=2 \sqrt{3 \pi
  M}$. Unfortunately our Anderson-Higgs mass term is not isotropic in
space-time, therefore to apply Eq.~\eqref{klei} we symmetrize it, and
{\it a posteriori} the precise choice of the coefficient turns out to be
almost irrelevant; we choose $\beta=|\Delta^s_0|^2/(12 \pi M)$. With
this choice the solution of Eq.~\eqref{klei} gives
\begin{eqnarray}
\label{Tc}
(|\Delta^s_0|^2)_c \approx \frac{m_s^2}{2}-\frac{m_s^4}{128 \pi^2}
\end{eqnarray}
and the choice of the symmetrized coefficient changes only the
second almost irrelevant term. According to
Fig.~\ref{fig:phasediagram} one obtains for the SC state at $T=0$
a range of dopings from $\delta \approx 0.04$ to $\delta \approx
0.25$.  Tentatively extending the formula Eq.~\eqref{Tc} to finite
$T$, one obtains for the critical temperature $T_c$ the red dashed
line in Fig.~\ref{fig:phasediagram}. For the critical value of
$\Delta^s_0$ the value of $M$ is quite small, hence $q$ almost
vanishes and within this approximation the SC transition is
essentially of XY-type. This implies also that the gauge
contributions of holons which have been neglected above would be
actually self-consistently strongly suppressed. In general one can
see from Eq.~\eqref{eq:54} that in our approach a reduction of
$M$, and hence of spinon kinetic energy, implies a reduction of
the gauge fluctuations.  The scale of $T_c$ in our approach is
reduced w.r.t. a naive BCS value $\sim \Delta^h_0$ because a price
has to be paid to overcome the spinon repulsion, so that its scale
is essentially set by $\Delta^s_0$.

Let us now outline some physical consequences of our approach to
SC that are presently under further investigation:

1) In the SC state the gauge gap destroys the Reizer singularity
(see Eq.~\eqref{reizer}) which is responsible for the anomalous
$T$-dependent life-time of the magnon and electron resonances in
the PG normal state. Hence these resonances become sharper at the
SC transition. In turn, this improves the kinetic energy of the
hole. Therefore in our approach the SC transition from PG is
``kinetic energy driven'',\cite{marsiglio} as opposed to the
standard BCS ``potential energy driven''.
%\textcolor{red}{I DO'NT KNOW IF PUT
%IT IN THE TEXT IN A NOTE OR OMIT Besides the already quoted
%relation with Trugman's ideas this  phenomenon is vaguely
%reminiscent of a mechanism discussed  in \cite{marsiglio}[J.Hirsh
%and F. Marsiglio, Phys. Rev. B 62 (2000) 15131] but the framework
%is entirely different.}
The above feature is supported by some experiments on optical
conductivity \cite{deutscher2005,heumen2007} and, in particular, a
recent experiment on underdoped cuprates,\cite{giannetti} where one
finds an increase of the kinetic energy in PG, being consistent with our
approach (due to the partial gap induced by the $\pi$-flux), and its
sharp decrease in the SC phase.  This shows that within this gauge
approach the compositeness of the hole, with a gauge gluing force coming
from the single-occupancy constraint, proved to be essential in
interpreting the transport\cite{marchetti2007} and
thermodynamical\cite{marchetti2008} properties of cuprate
superconductors, turns out to be a key feature also for the SC
transition.

 2) The appearance of two
positive branches in the spinon dispersion relation for a suitable
spinon-antispinon attraction mediated by gauge
fluctuations (in particular those corresponding
to the ${\bf Z}_2$ subgroup left unbroken by the condensation of
the SC pairs) induces a similar structure for the magnon
dispersion around the AF wave vector,\cite{gam} reminiscent of
the hour-glass
 shape of spectrum found in neutron experiments.\cite{hour} Furthermore,
since the energy of the resonance is approximately twice the
spinon gap $J m_s\sim J (1-2\delta) |\delta\ln\delta|^{1/2}$, it
has a maximum in $\delta$ near the maximum of $\Delta^s_0$, and
through Eqs.~\eqref{klei} and ~\eqref{Tc}, it is naturally related
to $T_{c}$.  This appears as a key feature of our approach:
because of the intimate relation between
 short-range AFO  and the SC attraction, both coming from
 the same term in the representation
of the $t-J$ model Eq.~\eqref{eq:8} (the third term, see also
~\eqref{eq:9}), there is an intrinsic relation between the energy
of the magnon resonance and $T_c$. This feature qualitatively
agrees with experiments,\cite{gre} as shown in
Fig.~\ref{fig:resonance}.

\section{Discussions and Conclusions}
Before concluding, let us briefly comment on the comparison of the
present proposal with other models on SC mechanism in cuprates.

It is clear that our proposal differs in an essential way from the
traditional BCS-Eliashberg approach,\cite{sch} no matter whether the
electron-phonon interaction or the AF fluctuations serve as the pairing
glue, SC being there ``potential energy driven''.  SC
arises in our approach from PG exhibiting characteristic features of a
doped Mott insulator, such as small FS, hence from the physical point of
view this approach is an implementation of the basic ideas advocated by
P.W. Anderson, attributing SC to the strong correlation effects in doped
Mott insulators.\cite{and1,bas,and2} Furthermore, in our approach the
leading part of the original Heisenberg term is used to provide the AF
action for the spinons, by using the identity Eq.~\eqref{ide} (holding
for the bosonic spinons).  Only the subleading term proportional to the
holon-pair density is used to obtain the formation of a finite density
of RVB-pairs in Eq.~\eqref{eq:61}, so the derived SC can be viewed as
vaguely reminiscent of Laughlin's gossamer SC.\cite{lau1}

Our formalism shares some similarities with other approaches,
exploring the same underlying physical idea, with, however, some
substantial differences. Both in the standard
slave-boson\cite{leeRMP} and in the bosonic-RVB
phase-string\cite{weng1,weng2} approaches the Nernst effect and SC
occur due to Bose-Einstein condensation (BEC) of bosonic holons.
Since BEC persists for arbitrary small density in these approaches
both Nernst effect and SC at $T=0$ occur as soon as the long-range
AFO disappears.
%In the new version\cite{weng10} of the bosonic RVB
%phase-string model at $T=0$ a finite interval opens up between the
%long-range AFO and SC due to the compact nature of the gauge
%fields; in this region, however, holons and spinons are
%``condensed'' in contrast to our approach.
The same also happens in the standard ``preformed pair''
approaches,\cite{ek} due to the persistence of condensation of
pairs in the extreme BEC limit. Instead, in our approach the
repulsive interaction between spinons prevents the appearance of
the Nernst effect below a critical doping, and the hole pairing
occurs only when the holon pair density is sufficiently large to
``force'' the RVB spinon pairing via gauge coupling, while an even
higher doping at $T=0$ is necessary to get SC. Similar
``critical'' dopings also appear in the phase-fluctuation approach
of Ref.~\onlinecite{tesanovich2008}, of which the main physical
difference from ours is in that approach nodes appear in the
Nernst phase, whereas in ours a finite FS still persists and nodes
appear only in the SC phase. Also, in the new version\cite{weng10}
of the bosonic RVB phase-string model at $T=0$ a finite interval
opens up between the long-range AFO and SC, due to the compact
nature of the gauge fields; in this region, however, holons and
spinons are ``condensed'' in contrast to our approach.

Another peculiar feature of the approach presented here,
distinctive from other approaches  is the appearance of three
distinct crossovers related to the PG phenomenology: in our
notations $T_{ph}$, $T_{ps}$ and $T^*$. The highest one in $T$ is
$T_{ph}$ (the presence of $t'$ there is relevant) where holons
start to pair reducing the spectral weight of the hole\cite{gam}
and producing, {\it e.g.}, a deviation from linear in-plane
resistivity.  A lower one, $T_{ps}$ where incoherent hole pairs
are formed, mainly affecting the magnetic properties since a
finite FS still persists, {\it e.g.}, giving rise to a boundary of
the diamagnetic/Nernst signal. Finally we have the crossover line
$T^*$ crossing $T_{ps}$ in the phase diagram; it is due to the
peculiar phenomenon of the optimal $\pi$-flux occurring only in
bipartite lattices and it is not directly related to SC. It
corresponds to a change in the holon dispersion and is
characterized by complete suppression of the spectral weight for
holes in the antinodal region.  Below $T^*$ the effect of
short-range AF fluctuations become stronger and their interplay
with thermal diffusion induced by gauge fluctuations gives rise to
the metal-insulator crossover and the inflection point of in-plane
resistivity.  Such composite structure of crossovers seems also to
emerge from recent experiments on optical
conductivity.\cite{giannetti2}

The relation found between $T_c$ and the energy of the magnetic
resonance might suggest that perhaps in some form at least part of the
mechanism for SC presented here can apply also to SC materials different
from cuprates, but with strong interplay between SC and AF. One possible
candidate is the recently discovered iron-arsenic
superconductors,\cite{pnic} which show similar phase diagrams as
cuprates. However, the parent compounds in those systems are not
insulators, but rather semi-metals. On the other hand, the freshly found
new iron-selenic systems\cite{fese} do have insulating states as
reference, and we can expect similar behavior to occur there.

To conclude, in this paper the spin-charge gauge approach is applied to
derive superconducting properties from the $t-J$ model with
single-occupancy constraint describing the Cu-O plane of underdoped
cuprate superconductors with the following distinct features:

1) The same model and the same set of approximations are used to
consider both normal and superconducting state properties without extra
assumptions. The physical implications of the theory are derived
explicitly, and in its totality are consistent with experimental
observations.

2) The interplay of antiferromagnetism and superconductivity is taken as
the underlying physical foundation and is implemented systematically for
both normal and superconducting states. The same super-exchange term is
giving rise to antiferromagnetism in the leading order, and producing
superconducting pairing in the sub-leading order. As a consequence, a
universal relation between the superconducting transition temperature
and the magnetic resonance mode energy is derived, in consistency with
experiments.

3) An unusual three-step scenario for the appearance of
superconductivity is proposed: At the higher crossover temperature
the charge carriers (holons) start to form pairs and they affect
the charge transport properties (deviation from the linear
temperature dependence of resistivity); at the lower crossover
temperature incoherent (local) hole pairs are formed, and the
derived pairing amplitude as a function of temperature/doping is
consistent with the Nernst, diamagnetism and STM data; the true
superconducting transition is derived as ``almost'' of the
classical 3D XY-type, with a phase diagram in agreement with
experiments.

\begin{acknowledgments}
  P.A. Marchetti acknowledges financial support of INFN, F. Ye is
  supported by NSFC Grant No. 10904081, and L. Yu thanks the NSF of
  China for the financial support. We acknowledge helpful discussions
  with A. Di Giacomo, C. Giannetti, Z. Tesanovic, Y. Y. Wang, H. H. Wen,
  and Z. Y. Weng.
\end{acknowledgments}

\appendix
\section{Diagonalization of Mean Field Hamiltonian of holon pairing}
\label{sec:diag-mf-hamilt}
We introduce a four components spinor field, $
\hat{\Psi}_{\alpha,\vec{k}}= (\hat{a}_{\alpha,\vec{k}},
\hat{b}_{\alpha,\vec{k}}, \hat{a}^{\dagger}_{\alpha,-\vec{k}},
\hat{b}^{\dagger}_{\alpha,-\vec{k}} )^{t} $, in terms of which the holon
Hamiltonian $\hat{H}_{h,\alpha} =\sum_{\vec{k}}
\hat{\Psi}^{\dagger}_{\alpha,\vec{k}} \mathcal{H}_{\alpha,\vec{k}}
\hat{\Psi}_{\alpha,\vec{k}}$ with the $4\times 4$ matrix
$\mathcal{H}_{\alpha,\vec{k}}$
\begin{eqnarray}
\label{eq:66}
\mathcal{H}_{\vec{k}} = \left(
    \begin{array}{cccc}
      -\mu  & v_Fk & 0&  \Delta^h_{\vec{k}} \\
      v_Fk  & -\mu  & - \Delta^h_{-\vec{k}} & 0\\
      0  &- \Delta^{h*}_{-\vec{k}}  & \mu  & -v_Fk \\
      \Delta^{h*}_{\vec{k}} &0 & -v_Fk &\mu
    \end{array}
  \right).
\end{eqnarray}
For the sake of simplicity, we omit temporarily the subscript $\alpha$.
One can introduce a unitary matrix $\mathcal{A}$,
\begin{eqnarray*}
  \mathcal{A} & = &\frac{1}{\sqrt{2}} \begin{pmatrix}
    1 & 1    & 0 & 0 \\
    -1 & 1 & 0 & 0 \\
    0 & 0& 1 & 1 \\
    0&0  & 1 & -1
  \end{pmatrix}
\end{eqnarray*}
which transforms the matrix $\mathcal{H}_{\vec{k}}$ to
\begin{eqnarray*}
  \mathcal{A}^{\dagger}\mathcal{H}_{\vec{k}}\mathcal{A}  =  \begin{pmatrix}
    -\mu-v_Fk & 0 & 0 & -\Delta^{h}_{\vec{k}} \\
    0 & -\mu+v_Fk & \Delta^{h}_{\vec{k}}  & 0 \\
    0 &\Delta^{h*}_{\vec{k}} & \mu- v_Fk &0 \\
    -\Delta^{h*}_{\vec{k}} &0& 0&\mu+v_Fk
  \end{pmatrix},
\end{eqnarray*}
provided that the holon pairing parameter $\Delta^{h}_{\vec{k}}$ is
p-wave like, i.e., $\Delta^h_{-\vec{k}} = -\Delta^h_{\vec{k}}$.  The
spectrum of quasiparticles consists of two decoupled branches,
\begin{eqnarray}
\label{eq:67}
\epsilon_{\pm,\vec{k}}^h = \sqrt{(v_Fk\pm \mu)^2+|\Delta^h_{\vec{k}}|^2}.
\end{eqnarray}
The free energy at temperature $T$ then reads
\begin{eqnarray}
\label{eq:68}
F=-T\sum_{i=\pm, \vec{k}} \left[
  \ln(1+e^{-\frac{\epsilon_{i \vec{k}}}{T}})+\ln(1+
  e^{\frac{\epsilon_{i \vec{k}}}{T}}) \right].
\end{eqnarray}
According to Hellman-Feynman theorem, we have the gap equation for order
parameter $\Delta_{\vec{q}}^h$,
\begin{eqnarray}
\label{eq:69}
&&\langle \hat{b}_{-\vec{q}}\hat{a}_{\vec{q}} \rangle = \frac{\partial F}{\partial
  \Delta_{\vec{q}}^{h*}} =-\sum_{i=\pm}
\frac{\Delta_{\vec{q}}^{h}}{2 \epsilon_{i,\vec{q}}^h}\tanh \left(
  \frac{\epsilon_{i,\vec{q}}}{2T} \right) \nonumber\\
&&\Delta_{\vec{k}}^h=\sum_{\vec{q}}V_{\text{eff}}(\vec{k}-\vec{q})
\langle \hat{b}_{-\vec{q}}\hat{a}_{\vec{q}} \rangle.
\end{eqnarray}
If we assume $\mu>0$, the branch with energy $\epsilon_{-,\vec{k}}^h$(as
given in Eq.~\eqref{eq:25}) is lower and responsible for the low energy
physics of $p$-wave pairing.  The corresponding quasiparticle field
reads
\begin{eqnarray}
\label{eq:70}
\hat{\psi}_{\vec{k}} =
 \frac{1}{\sqrt{2}} (\hat{a}_{\vec{k}} + \hat{b}_{\vec{k}}).
\end{eqnarray}
In terms of $\hat{\psi}$-fields, the effective pairing Hamiltonian can
be written as
\begin{eqnarray}
\label{eq:71}
\hat{H}_{\text{eff}}^h = \sum_{\vec{k}}(v_Fk-\mu)\hat{\psi}^{\dagger}_{\vec{k}}
\hat{\psi}_{\vec{k}}  -\frac{1}{2}(\Delta^{h}_{\vec{k}}
  \hat{\psi}^{\dagger}_{\vec{k}} \hat{\psi}^{\dagger}_{-\vec{k}} + h.c.),
\end{eqnarray}
and the gap equation at temperature $T$ can be obtained by neglecting
the positive branch with $i=+$ in Eq.~\eqref{eq:69} as already written
in ~\eqref{eq:24}.

For the right Dirac cone, the p-wave pairing parameter takes the
following form in polar coordinate system
\begin{eqnarray}
\label{eq:72}
\Delta_{\vec{q}}^h = \Delta^h_0(q)(\cos\theta_{\vec{q}}-\sin
\theta_{\vec{q}})
\end{eqnarray}
with its angular and radial parts separated consistently with the
gap equation. Substituting Eq.~\eqref{eq:72} into Eq.~\eqref{eq:24}, we
have at zero temperature
%\begin{widetext}
\begin{eqnarray}
\label{eq:73}
&&\Delta^h_{\vec{k}} =  \int \frac{d^2\vec{q}}{8\pi^2} \nonumber\\
&&\frac{J_{\text{eff}}\times 2kq\cos(\theta_{\vec{k}}-\theta_{\vec{q}})}
{(k^2-q^2)^2+\ell_s^{-4}+2(k^2+q^2)\ell_s^{-2}+4k^2q^2\sin^2(\theta_{\vec{k}}-\theta_{\vec{q}})} \nonumber\\
&&\times\frac{\Delta^h_0(q)(\cos\theta_{\vec{q}}-\cos\theta_{\vec{q}})}{
  \sqrt{(\mu-v_Fq)^2+[\Delta^h_0(q)(\cos\theta_{\vec{q}}-\cos\theta_{\vec{q}})]^2}}
\end{eqnarray}
%\end{widetext}
Note that the most important term comes from momentum around the Fermi
surface,  $q\sim k$ and $\theta_{\vec{q}}\sim \theta_{\vec{k}}$ or
$\theta_{\vec{q}}\sim \pi + \theta_{\vec{k}}$, therefore we can neglect
terms proportional to $\sin^2(\theta_{\vec{k}}-\theta_{\vec{q}})$ in the
denominator of the first fraction of the r.h.s. of
Eq.~\eqref{eq:73}. Then the angular part is just dropped off from the
p-wave gap equation with only the radial part remaining
\begin{eqnarray}
\label{eq:74}
\frac{\Delta_{0}^h(k)}{J_{\text{eff}}} = k\int_0^{\Lambda}
\frac{dq}{8\pi^2}\frac{q^2\times G \left[ \frac{\mu-v_Fq}{\Delta_0^h(q)}
  \right] }{(k^2-q^2)^2+\ell_s^{-4}+2\ell_s^{-2}(k^2+q^2)}
\end{eqnarray}
with
\begin{eqnarray}
\label{eq:75}
G(x)&\equiv&\int_0^{2\pi}d\theta \frac{1-\sin(2\theta)}{\sqrt{x^2
    +[1-\sin(2\theta)]^2}} \nonumber\\
&=& 4x \left[ E \left( -\frac{2}{x^2} \right) - K \left( -\frac{2}{x^2} \right) \right]
\end{eqnarray}
where $\Lambda$ is a momentum cutoff and $E(x)$ and $K(x)$ are the
elliptic integral of first and second kink, respectively.

\section{Evaluation of the partition function of vector bosons}
\label{sec:eval-part-funct}

In this appendix, we show how to compute the path integral of
Eq.~\eqref{eq:55} of vector bosons including the contribution from
the holon part(see Eq.~\eqref{eq:37}). The relevant Lagrangian can be
decomposed into two parts
\begin{eqnarray}
\label{eq:76}
\mathscr{L}_g = \frac{1}{3\pi M} [a_{\mu} (-\partial^2g^{\mu\nu}
+ \partial_{\mu}\partial_{\nu}) a_{\nu} + a_{\mu} m^{\mu\nu} a_{\nu} ]
\end{eqnarray}
where the holon's contribution is absorbed into mass term $m^{\mu\nu}$
which in momentum space has the form
\begin{eqnarray}
\label{eq:77}
m^{\mu\nu} = \begin{pmatrix}
|\Delta_0^s|^{2}+\tilde{c}_2 & 0 & 0 \\
0 & \frac{|\Delta_0^s|^2}{2} + f(\vec{k})  & 0 \\
0 & 0 & \frac{|\Delta_0^s|^2}{2} + f(\vec{k})
\end{pmatrix}.
\end{eqnarray}
with $\tilde{c}_2=3\pi Mc_2$ and $f(\vec{k})=3\pi M c_{1}
\sqrt{v_F^2k_{+}^2 + v_{\Delta}^2k_{-}^2}$(see Eq.~\eqref{eq:28}).
%with $\tilde{c}_{1,2} = 3\pi M c_{1,2}$.

It is not appropriate to integrate $a_\mu$ directly, because the
redundancy of degree of freedom of vector bosons. To rule out the
redundancy, we adapt a method developed by 't Hooft.\cite{tH} We first
rewrite the $a$-field in terms of the h/s gauge field $A_{\mu}$ and the
phase field $\phi^h$:
\begin{eqnarray}
\label{eq:78}
a_{\mu} = A_{\mu} -\frac{1}{2}\partial_{\mu} \phi^{h} .
\end{eqnarray}
Clearly, $a_{\mu}$ is gauge invariant under the following h/s gauge
transformation
\begin{eqnarray}
\label{eq:79}
&&A_{\mu} \rightarrow A_{\mu} + \partial_{\mu}\Lambda \nonumber\\
&&\phi^h \rightarrow \phi^h + 2 \Lambda
\end{eqnarray}
The Lagrangian rewritten in terms of $A_{\mu}$ and $\phi^h$ is given by:
\begin{eqnarray}
\label{eq:80}
3\pi
M\mathcal{L}_{g}&=&A_{\mu}(-\partial^2g^{\mu\nu}+\partial^{\mu}\partial^{\nu}+
m^{\mu\nu})A_{\nu} \nonumber\\
&&-\frac{1}{4}\phi^{h} \partial_{\mu}m^{\mu\nu}\partial_{\nu}\phi^h
 +\phi^h m^{\mu\nu}  \partial_{\mu}A_{\nu}
\end{eqnarray}
We choose the fixing gauge function as
\begin{eqnarray}
\label{eq:81}
F = -{m^{\mu\nu}}\partial_{\mu}A_{\nu} + \frac{1}{2} \phi^{h}
\end{eqnarray}
whose derivative with respect to the infinitesimal gauge transformation
Eq.~\eqref{eq:78} reads
\begin{eqnarray}
\label{eq:82}
\frac{\delta F}{\delta \Lambda}= -m^{\mu\nu} \partial_{\mu}\partial_{\nu} + 1\equiv D.
\end{eqnarray}
% The arbitrary function $\sigma(x)$ is irrelevant to the final results,
% and can be eliminated by integrated out with the Gaussian weight factor
% \begin{eqnarray*}
% \frac{1}{\Omega}e^{- \frac{1}{3\pi M}\int d^3x\sigma^2(x)}
% \end{eqnarray*}
% where $\Omega$ is a normalization factor.
Then, following the Fadeev-Popov-Dewitt's approach, the path integral
involving only the physical degrees of freedom can be calculated as
\begin{eqnarray}
\label{eq:83}
Z_{g}&=&
%\frac{1}{\Omega}
\int D[A_{\mu},\phi^h] \left| \frac{\delta F}{\delta \Lambda} \right|
e^{- \int d^3x\left[\frac{1}{3\pi M}A_{\mu}K^{\mu\nu}A_{\nu}+ \frac{1}{4}\phi^h D\phi^h \right] } \nonumber\\
&=&\det[m^{00}(3\pi M)^{3/2} D^{1/2}K^{-1/2}]
\end{eqnarray}
with
\begin{eqnarray}
\label{eq:84}
K= -\partial^2g^{\mu\nu} +\partial^{\mu}\partial^{\nu} +m^{\mu\nu}
 - {m^{\mu\mu'}m^{\nu\nu'}}\partial_{\mu'}\partial_{\nu'}
\end{eqnarray}
Note that $m^{00}$ in the determinant of Eq.~\eqref{eq:83} is originated
from the complete measure of $D[\Delta_e^{*},\Delta_e]$ of
Hubbard-Stratonovich transformation.  Finally, we obtain the following
result
\begin{eqnarray}
\label{eq:85}
Z_g = \prod_{\omega,\vec{k}} \frac{(3\pi M)^{\frac{3}{2}}}{(\omega^2 + \vec{k}^2 +
   m^{11} )^{\frac{1}{2}} (\omega^2 +
   m^{11} + \frac{m^{11}}{m^{00}} \vec{k}^2 )^{\frac{1}{2}}}
\end{eqnarray}
which is given in Sec.~\ref{sec:gap-equation-spinon}.  The poles of
Eq.~\eqref{eq:85} lead to the spectra of gauge bosons in
Eq.~\eqref{eq:64}.

\section{Coherence of holon pairs and gapless gauge field}
\label{sec:hol-p-coh}
 We sketch here the argument proving that the gapless
transverse gauge field arising from Eq.~\eqref{eq:37} is
inconsistent with the coherence of holon pairs in PG, i.e. with a
non-vanishing expectation value of $e^{i \phi^h}$ at $T=0$ in the
Coulomb gauge, implying that the (global) h/s symmetry is broken.
Let us assume condensation of holon pairs, but not of spinon
pairs, then the Anderson-Higgs ``mass'' term in Eq.~\eqref{eq:54}
at large scale simply renormalize the Maxwell term.\cite{seiler}
Then in the Coulomb gauge the effective lagrangian for $A_\mu$ and
$\phi^h$ has the following form:
\begin{eqnarray}
\label{eq:86}
\mathscr{L}[A_\mu,\phi^h](x)=[c_0(A_i(\Delta+\partial_0^2)A_i+A_0\Delta A_0)\nonumber\\
+c_1 (A_i \sqrt{\Delta} A_i +\partial_i\phi^h\sqrt{\Delta} \partial_i\phi^h)\nonumber\\+
c_2(A_0+\partial_0\phi^h)^2](x)
\end{eqnarray}
with $c_i,i=0,1,2$ suitable positive constants.  Integrating out the
gauge field in the path-integral formalism we obtain the effective
action for $\phi^h$ in momentum space
\begin{eqnarray}
\label{eq:87}
\mathscr{L}[\phi^h](\vec{k},\omega)=\phi^h(\vec{k},\omega)[c_1 |\vec{k}|^{3}\nonumber\\
+(c_2^{-1}+(c_0|\vec{k}|^2)^{-1})^{-1} \omega^2]\phi^h(-\vec{k},-\omega).
\end{eqnarray}
Neglecting the subleading $c_2$ term one can easily calculate the
equal-time large-distance behaviour of the Green function of $\phi^h$:
\begin{eqnarray}
\label{eq:88}
G(\vec{x}, x^0=0)=\int d^2k d\omega \frac{e^{i \vec{k}\cdot
    \vec{x}}}{c_1 |\vec{k}|^{3}+c_0|\vec{k}|^2 \omega^2} \sim
|\vec{x}|^{1/2}.
\end{eqnarray}
Therefore we have for large $|\vec{x}|$:
\begin{eqnarray}
\label{eq:89}
\langle e^{i \phi^h(\vec{x},0)}e^{-i \phi^h(0)} \rangle \sim e^{-c|\vec{x}|^{1/2}},
\end{eqnarray}
vanishing at large distance, so that the condensation cannot occur.

%\bibliographystyle{apsrev4-1}
%\bibliography{Superc}

\end{document}